\documentstyle[aps,preprint,prb,tighten]{revtex}
\begin{document}
\draft
\title{ Signatures of Chaos in the Statistical Distribution of
       Conductance Peaks in Quantum Dots}

\author{ Y. Alhassid$^1$ and
	 C. H. Lewenkopf$^2$}

\address{$^1$ Center for Theoretical Physics, Sloane Physics Laboratory,
	 Yale University, New Haven, Connecticut  06520, USA\\
	 $^2$ Instituto de F\'{\i}sica, Universidade de S\~ao Paulo,
	 C.P. 66318, 05389-970 S\~ao Paulo, Brazil}

\date {Submitted May 1, 1996}
\maketitle

\begin{abstract}
Analytical expressions for the width and conductance peak
distributions of irregularly shaped quantum dots in the
Coulomb blockade regime are presented in the limits of 
conserved and broken time-reversal symmetry.
The results are obtained using random matrix theory and are
valid in general for any number of non-equivalent and correlated
channels, assuming that the underlying classical
dynamic of the electrons in the dot is chaotic or that the dot is
weakly disordered.
The results are expressed in terms of the channel correlation
matrix  which for chaotic systems is given in closed form for both
point-like contacts and  extended leads.
We study the dependence of the distributions on the number of channels
and their correlations.
The theoretical distributions are in good agreement with those
computed  in a dynamical model of a chaotic billiard.
\end{abstract}

\narrowtext

\section{Introduction}

One of the most interesting aspects of electron transport in submicron
scale devices is the interplay between quantum coherence and aperiodic
but reproducible conductance fluctuations.
Over the past decade the phenomenon of universal conductance fluctuations
in disordered systems (where impurity scattering dominates) 
has been understood through the use of stochastic models.
 More recently, a new generation of experiments \cite{Marcus92} was
designed to measure conductance fluctuations in the ballistic regime
where the dynamics of the electrons in the device is determined by the
geometry of its boundary.
The stochastic approach to these systems is justified by the underlying
classical chaotic dynamics.
This situation is distinct from the diffusive case, where the
corresponding classical limit of the quantum problem  is not
fully understood.

In this paper we discuss the conductance fluctuations in quantum dots.
\cite{Kastner92,Meirav90,Foxman93,Foxman94} 
These are semiconductor devices in which the electrons
 are confined to a  two-dimensional 
region whose typical linear dimension is in the submicron range.
In particular we are interested in the Coulomb blockade regime
where the leads are weakly coupled to the dot, either
because the leads are very narrow, or due to the presence of
potential barriers at the lead-dot interface. \cite{Kastner92}
The electrons inside the dot are characterized by isolated resonances
whose width is smaller than their average spacing, and conductance
occurs through resonant tunneling.
As a consequence, the conductance peaks when the Fermi energy matches
a resonance energy of the electrons inside the dot and an additional
electron tunnels into the dot.
Such a system resembles the compound nucleus in its region of isolated
resonances. \cite{Porter65}
The macroscopic charging energy required to add an electron to a dot
is determined by its capacitance $C$ and is given by $e^2/C$.
Since $C$ is a constant which is determined essentially by
 the geometry of the dot,  the conductance
  exhibits equally spaced oscillations as a function of the gate voltage
 (or Fermi energy).
At low temperatures $\Gamma \ll kT < \Delta$ the width of the conductance
peaks is $\sim kT$, but the heights exhibit order of magnitude variations.
\cite{Meirav90,Foxman93,Foxman94}

When the electron-impurity mean free path is larger than the size of the dot,
the classical dynamics of the electron inside the dot
 is determined by the scattering from the dot's boundary.
 Due to small irregularities in the dot's shape, the electron
 displays chaotic motion, and its quantum transport
through the dot can be described by  statistical 
$S$-matrix theory.\cite{Lewenkopf91}
Since the Coulomb blockade regime is dominated by resonances,
the conductance peaks can be used to probe the chaoticity 
 of the underlying  resonance wavefunctions.
A statistical theory of the conductance peaks was originally developed
in Ref. \onlinecite{Jalabert92}.
By using $R$-matrix theory, \cite{Wigner47,Lane58}  the
conductance peak amplitude was expressed in terms of the electronic
resonance wavefunction across the contact region between the
 dot and the leads.
When the dynamics of the electron inside the dot is chaotic,  
the fluctuations of the wavefunction inside the dot are assumed to be well
described  by random matrix theory (RMT).
In Ref.\onlinecite{Jalabert92} the conductance distribution was derived
in closed form for one-channel leads.
These results were rederived in Ref. \onlinecite{Prigodin93}, and
later extended to the case of two-channel leads in the absence of
time-reversal symmetry \cite{Mucciolo95} through the use of
the supersymmetry technique.\cite{Efetov83}
However,  the calculations required by this technique become
 too complicated 
to apply in the general case of any
number of possibly correlated and/or non-equivalent channels.

The conductance distributions for one-channel leads were recently
measured \cite{Chang96,Folk96} and found to be in agreement with
theory for both  cases of conserved and broken time-reversal symmetry.
This indicates that the dephasing effect, which plays an important role
in open dots, \cite{Chan95,Baranger95} is of little importance
for closed dots.

In this paper we discuss in detail  the width and conductance peak
distributions for leads with any number of channels that are in
general correlated and non-equivalent.
Exact closed expressions for these distributions are derived for
both cases of conserved and broken time-reversal symmetry.
\cite{Alhassid95}
We find that these distributions are entirely characterized by the
eigenvalues of the channel correlation matrices $M^l$ and $M^r$ in
the left and right leads, respectively.
The strength of our approach is in its simplicity, since it relies
solely on standard RMT techniques.
To test our predictions we compare our analytical findings to
numerical simulations of a chaotic dynamical model, the conformal
billiard. \cite{Robnik83}
Statistical width and conductance distributions of one-channels leads
were recently studied in detail in this model. \cite{Bruus94}
Although our paper deals mainly with ballistic dots whose classical
dynamics is chaotic, our results should also be valid in the diffusive
regime of weakly disordered dots, where random matrix theory is
applicable.

We note  that under certain conditions the partial width is
analogous to the wavefunction intensity at a given point.
Therefore our width distributions can also be tested by microwave
cavity experiments,\cite{Alt95,Sridhar91} where the
 intensities are measured
at several points that are spatially correlated.

The plan of the paper is as follows: In Section II we
briefly review  the conductance in quantum dots in their
Coulomb blockade regime.
In Section III we discuss the statistical model and derive analytic
results for the partial and total width distributions in each lead,
 for the channel
correlation matrix and for the conductance distribution. We
investigate the variation of these distributions as a function
of the number of channels and their sensitivity to the degree
of correlations between them.
Those findings are compared in Section IV with numerical results
obtained for the conformal billiard.
Finally, in Section V we discuss the validity of our assumptions in the
the context of typical experiments.

\section{Conductance in Quantum Dots}

In this section we briefly review the formalism and introduce the
notation used throughout this paper.
In particular, we express the conductance peak heights in terms of
the channel and resonance wavefunctions of the dot.

For $\Gamma \ll kT \ll \Delta$, which is typical of many experiments,
\cite{Meirav90} the observed on resonance conductance peak
amplitude is given by \cite{Beenakker91,Zirnbauer93}

\begin{eqnarray}
\label{condpeak}
    G_\lambda =\frac{e^2}{h}\, \frac{\pi}{2 kT} \, g_\lambda
\qquad \mbox{with} \qquad
    g_\lambda =\frac{\Gamma^l_\lambda \Gamma_\lambda^{r}}
	    {\Gamma_\lambda^{l}+\Gamma_\lambda^{r}} \;,
\end{eqnarray}
where $\Gamma_\lambda^{l(r)}$ is the partial decay width
of the resonance $\lambda$ into the left (right) lead.
Since each lead can support several open channels we have
$\Gamma_\lambda^{l(r)}=\sum_c\Gamma_{c\lambda}^{l(r)}$,
where $\Gamma_{c\lambda}^{l(r)}$ is the partial width to decay
into  channel $c$ in the left (right) lead.

In the $R$-matrix formalism, \cite{Lane58} the partial widths are related to
 the resonance wavefunction inside the dot.
More specifically, introducing the partial amplitudes $\gamma_{c\lambda}$,
such that $\Gamma_{c\lambda}= |\gamma_{c\lambda}|^2$, one can write
\begin{eqnarray}
\label{gamdefR}
\gamma_{c\lambda} = \sqrt{\frac{\hbar^2 k_c P_c}{m} }
     \int \! dS\,\Phi_c^*(\bbox{r}) \Psi_\lambda(\bbox{r})  \;.
\end{eqnarray}
Here $\Psi_\lambda(\bbox{r})$ is the $\lambda$--th  resonance
wavefunction in the dot, $\Phi_c(\bbox{r})$ is the transverse
wavefunction in the lead that corresponds to an open channel $c$,
and the integral is taken over the contact area between the lead
and the dot.
$k_c$ and $P_c$ are the longitudinal wavenumber and penetration factor
in channel $c$, respectively.

Eq.~(\ref{gamdefR}) shows that the contributions to the partial width
amplitude from the internal and external regions of the dot factorize.
The information from the region external to the dot is contained in
$k_c$ and $P_c$.
These quantities are determined by the wave dynamics in the leads and
are non-universal.
They affect the average widths and enter explicitly in the correlation
matrix $M$.
However, the fluctuation properties of the conductance are generic
and depend only on the statistical properties of the electronic
wavefunction at the dot-lead boundary inside the barrier region.

A different physical modeling of a quantum dot assumes point-like
contacts and each lead is composed of several such point contacts.
\cite{Prigodin93,Mucciolo95}
In this model the conductance is also given by (\ref{gamdefR}) with
each point contact $\bbox{r}_c$ considered as one channel.
The corresponding  partial width is \cite{Prigodin93}
\begin{eqnarray}
\label{3}
\gamma_{c\lambda} = \sqrt{\frac{ \alpha_c {\cal A} \Delta}{\pi}}\,
	 \Psi_\lambda(\bbox{r}_c) \;,
\end{eqnarray}
where  ${\cal A}$ is the area of the dot, $\Delta$ is the mean spacing
and $\alpha_c$ is a  dot-lead coupling parameter.

Both models can be treated by our formalism.
This becomes apparent after the following considerations.
A resonance eigenfunction with eigenenergy $E=E_\lambda$ can be
approximated by an expansion in a fixed basis $\rho_\mu$
 of wavefunctions
 with the given energy $E$ inside the dot
\begin{eqnarray}
\label{4}
 \Psi_\lambda({\bf r})=
      \sum_\mu  \psi_{\lambda\mu} \,\rho_\mu(\bbox{r}) \;.
\end{eqnarray}
The sum over $\mu$ is truncated at $N$  basis states, where
$N$ is large and determined by precision requirements.
The  partial width in channel $c$ can then be expressed by the
scalar product
\begin{eqnarray}
\label{5}
 \gamma_{c\lambda} = \langle \bbox{\phi}_c | \bbox{\psi}_\lambda \rangle
     \equiv \sum_\mu \phi_{c\mu}^* \psi_{\lambda\mu} \;,
\end{eqnarray}
where
\begin{eqnarray}
\label{5'}
  \phi_{c\mu} \equiv \sqrt{\frac{\hbar^2 k_c P_c}{m} }
       \int\! dS \, \Phi_c^*(\bbox{r})\,  \rho_\mu (\bbox{r})
\end{eqnarray}
for the extended leads model, and
\begin{eqnarray}
\label{5''}
  \phi_{c\mu} \equiv  \sqrt{{ \alpha_c {\cal A} \Delta \over \pi}}
 \rho^*_\mu (\bbox{r}_c)
\end{eqnarray}
for the point contact model.
Thus, we are led to similar formulations of both the extended leads
and point-like contacts problems;  in the corresponding $N$-dimensional
space the partial width amplitudes of a level are simply the projections of its
corresponding eigenstate vector $\bbox{\psi}_\lambda$ on the channel vectors
$\bbox{\phi}_c$.
 The only difference between the two models is the explicit expression
for the channel vector $\bbox{\phi}_c$.
We note that the scalar product (\ref{5}) (that will be used
throughout this paper) is different from the original scalar product
defined in the spatial region extended by the dot.

\section{Statistical Model}

Due to the irregularity of the dot's shape, the motion of the electron
inside the dot is expected to be chaotic.
In Ref. \onlinecite{Jalabert92} we have developed a statistical theory
of the conductance peaks by assuming that the vectors $\psi_\lambda$
that correspond to the resonance wavefunctions inside the dot have
the same statistical properties as the eigenvectors of a random matrix
ensemble.
Here we study the limits of conserved time-reversal symmetry,
corresponding to the Gaussian Orthogonal Ensemble (GOE), and of
broken time-reversal symmetry, corresponding to the Gaussian
Unitary Ensemble (GUE).
The transition from one symmetry to another occurs when an external
magnetic field is applied.
The width distribution (or equivalently the wavefunction intensity
distribution) was derived in the crossover regime between 
symmetries  for the case of one channel leads only. \cite{Sommers94}

\subsection{The Joint Distribution of Partial Width Amplitudes}

In RMT the eigenvector $\bbox{\psi} \equiv (\psi_1, \psi_2, \ldots,
\psi_N)$ (here and in the following we omit the eigenvector label
$\lambda$)  is distributed randomly \cite{Brody81} on a sphere
$P(\bbox{\psi}) \propto \delta (\sum_{\mu=1}^N | \psi_\mu|^2-1)$.
The joint distribution of the partial width amplitudes $\bbox{\gamma} =
(\gamma_1, \gamma_2, \ldots, \gamma_\Lambda)$ for $\Lambda$
channels is then given by
\begin{eqnarray}
\label{6}
  P(\bbox{\gamma}) = \frac{\Gamma(\beta N/2)}{\pi^{\beta N/2}}
   \int \! D[\bbox{\psi}] \left[\prod_{c=1}^{\Lambda}
   \delta ( \gamma_c - \langle\bbox{\phi}_c|\bbox{\psi}\rangle) \right]
   \delta \!\left( \sum_{\mu=1}^N |\psi_\mu |^2 -1 \right) \;,
\end{eqnarray}
where $D[\bbox{\psi}] \equiv \prod_{\mu=1}^{N}{d\psi_\mu}$ for the GOE and
$D[\bbox{\psi}] \equiv \prod_{\mu=1}^{N}{d\psi^*_\mu d\psi_\mu/2\pi i}$
for the GUE.
To evaluate (\ref{6}) we transform the $\Lambda$ channels to a new set
of orthonormal channels $\hat{\bbox{\phi}}_c$
\begin{eqnarray}
\label{newbasis}
\bbox{\phi}_c =\sum_{c^\prime} \hat{\bbox{\phi}}_{c^\prime} F_{c^\prime c}
	      \qquad \mbox{with} \qquad
\langle \hat{\bbox{\phi}}_c | \hat{\bbox{\phi}}_{c^\prime} \rangle=
\delta_{cc^\prime} \;.
\end{eqnarray}
We then take advantage of the invariance of the corresponding Gaussian
ensemble under an orthogonal (unitary) transformation to rotate the
eigenvector $\bbox{\psi}$ such that its first $\Lambda$ components are
along the new orthonormal channels.
Denoting by ${\cal O}$ the orthogonal (unitary) matrix whose first
$\Lambda$ rows are the orthonormal vectors
$\hat{\phi}_c (c=1,\ldots, \Lambda)$, we change variables  in (\ref{5}) to
$\hat{\psi}_\mu =\sum_{\nu} {\cal O}_{\mu \nu}  \psi_{\nu}$.
Using  $\hat{\psi}_c=\langle\hat{\bbox{\phi}}_c |\bbox{\psi}\rangle$
we find

\begin{eqnarray}
\label{8}
  P(\bbox{\gamma}) = \frac{\Gamma (\beta N/2)}{\pi^{\beta N/2} }
     \int \!
  && \left( \prod^\Lambda_{c=1} d  \hat{\psi}_c \right)
     \left( \prod^N_{\mu=\Lambda+1} d \hat{\psi}_\mu \right)
     \left[ \prod^\Lambda_{c=1} \delta
   \left( \gamma_c - F_{c^\prime c} \hat{\psi}_{c^\prime} \right)
   \right] \nonumber \\
  && \times \, \delta\!\left( \sum^\Lambda_{c=1} |\hat{\psi}_c|^2 +
    \sum^N_{\mu=\Lambda+1} |\hat{\psi}_\mu|^2 -1 \right) \;.
\end{eqnarray}
The integration over these first $\Lambda$ components is now
easily done and gives

\begin{eqnarray}
\label{9}
 P(\bbox{\gamma}) = \frac{\Gamma(\beta N/2)}{\pi^{\beta N/2}\,|\det F|}
 \int\! D[\bbox{\hat\psi}]\,\delta \!\left(\bbox{\hat\gamma}^\dagger
				       \bbox{\hat\gamma} +
 \sum_{\mu=\Lambda+1}^N | {\hat\psi}_\mu |^2 -1 \right) \;,
\end{eqnarray}
where $\hat{\gamma}_c \equiv \langle\hat{\bbox{\phi}}_c |\bbox{\psi}\rangle =
\sum_{c^\prime} \gamma_{c^\prime}  F_{c^\prime c}^{-1*}$ are the
partial widths to decay to the new channels and the metric is
as before but excluding the first
$\Lambda$ components of $\bbox{\psi}$.
Finally, the latter integral is easily done by introducing spherical
coordinates in the $N-\Lambda$ dimensional space.
We obtain
\begin{eqnarray}
\label{10}
  P(\bbox{\gamma}) =
  \frac{\Gamma ( \beta N/2)}{\pi^{\beta \Lambda/2}
	\Gamma \left(\beta (N-\Lambda)/2\right) |\mbox{det} F|}\,
  \left[ 1 - \bbox{\gamma}^\dagger (F^\dagger F)^{-1} \bbox{\gamma} \right]
	^{\beta\frac{N-\Lambda}{2} -1} \;.
\end{eqnarray}

For $\Lambda \ll N$ and in the limit $N\rightarrow \infty$, we
recover a simplified expression
\begin{eqnarray}
\label{11}
 P(\bbox{\gamma}) = (\det M)^{-\beta/2}
	     \, \mbox{e}^{- \frac{\beta}{2}
	     \bbox{\gamma}^\dagger M^{-1} \bbox{\gamma} }\;,
\end{eqnarray}
where the matrix $M \equiv (NF^\dagger F)^{-1}$ is just the metric defined
by the original channels
\begin{eqnarray}
\label{12}
  M_{cc^\prime} = \frac{1}{N} \langle \bbox{\phi}_c |\bbox{\phi}_{c^\prime}\rangle \;.
\end{eqnarray}

The distribution (\ref{11}) is normalized with the measure
$D[\bbox{\gamma}] \equiv \prod_{c=1}^{\Lambda}{d\gamma_c/2\pi}$
for the GOE and $D[\bbox{\gamma}]\equiv\prod_{c=1}^{\Lambda}
{d\gamma^*_c d\gamma_c/2\pi i}$ for the GUE.
Note that for both ensembles the joint partial width amplitudes
distribution is Gaussian, the main difference being that the partial
amplitudes are real for the GOE and complex for the GUE.
Such a Gaussian distribution is also obtained by assuming that the
distribution is form-invariant under an orthogonal (unitary)
transformation. \cite{Krieger63}

It follows from (\ref{11}) that the matrix $M$ is just the correlation
matrix of the partial widths
\begin{eqnarray}
\label{13}
M_{cc^\prime} = \overline{ \gamma_c^* \gamma_{c^\prime}} \;.
\end{eqnarray}
In general the channels are  correlated and non-equivalent, {\sl i.e.}
non-equal average partial widths.
According to (\ref{12}) this is equivalent to assuming channels that are
non-orthogonal and have non-equal norms.

\subsection{The Channel Correlation Matrix $M$}

We shall now  derive explicit expressions for the correlation
matrix $M$ in a chaotic quantum dot.
Using Eq.~(\ref{12}) and  the definition of the scalar
product (\ref{5}) we find

\begin{eqnarray}
\label{14}
M_{cc^\prime}=\frac{\hbar^2}{2m} \sqrt{k_c k_{c^\prime} P_c P_{c^\prime}}
  \int\!dS\!\int\! dS^\prime \,\Phi^*_c(\bbox{r})
  \left[ \frac{1}{N} \sum_\mu \rho_\mu (\bbox{r})
  \rho_\mu (\bbox{r}^\prime) \right]
\Phi_c(\bbox{r}^\prime)\;.
\end{eqnarray}

We first discuss the case where there is no magnetic field so that the
motion inside the dot is that of a free particle.
Therefore, a resonance eigenstate inside the dot at energy $E
=\hbar^2k^2/2m$ can be expanded in a basis of free particle states
at the given energy $E$.
Since RMT is applicable on a local energy scale, this is the fixed basis
$\rho_\mu$ for which the eigenvector coefficients $\psi_\mu$ are
distributed randomly (on the sphere).
Using polar coordinates, such a basis of free waves is given by
$\rho_\mu (\bbox{r}) \propto J_\mu(kr)\exp(i\mu\theta)$ with
$\mu=0,\pm 1, \pm2,\ldots$, where $J_\mu$ are Bessel functions of
the first kind.
Denoting by $N$ the number of such waves on the energy shell, we find

\begin{eqnarray}
\label{15}
  \frac{1}{N} \sum_\mu \rho^*_\mu(\bbox{r}) \rho_\mu(\bbox{r}^\prime)=
  \frac{1}{{\cal A}}\sum_\mu J_\mu(kr) J_\mu(kr^\prime) 
{\rm e}^{i\mu(\theta^\prime - \theta)} =
  \frac{1}{{\cal A}}J_0(k| \bbox{r} - \bbox{r}^\prime |) \;,
\end{eqnarray}
where we have used the addition theorem for the Bessel functions.
\cite{Lebedev72}
A similar relation holds if we choose a plane waves basis
$\rho_\mu(\bbox{r}) = {\cal A}^{-1/2} \exp(i\bbox{k}_\mu\cdot\bbox{r})$
at a fixed energy $\hbar^2k^2/2m$ but with random orientation of
$\bbox{k}_\mu$ and use the integral representation of $J_0$.
With help of Eq.~(\ref{15}) we obtain for the correlation matrix

\begin{eqnarray}
\label{16}
  M_{cc^\prime} =
    \frac{\hbar^2}{2m {\cal A}}\sqrt{k_c k_{c^\prime} P_c P_{c^\prime}}
    \int \! dS \! \int \! dS^\prime \, \Phi^*_c(\bbox{r})
 J_0(k|\bbox{r} - \bbox{r}^\prime|) \Phi_c(\bbox{r}^\prime)\;,
\end{eqnarray}
for extended leads, while for the point contact model we find

\begin{eqnarray}
\label{17}
 M_{cc^\prime} = \frac{ \alpha_c \Delta}{\pi}
		 J_0(k| \bbox{r}_c - \bbox{r}_c^\prime|)\;.
\end{eqnarray}
We remark that Eq.~(\ref{17}) is equivalent to
$C(k|\Delta\bbox{r}|)  \equiv \overline{\Psi^*(
\bbox{r}) \Psi(\bbox{r}^\prime)}/\overline{|\Psi(\bbox{r})|^2} =
J_0(k|\bbox{r}-\bbox{r}^\prime|)$.
This result was first derived in Ref. \onlinecite{Berry77} based on
the assumption that the Wigner function of a classically chaotic
system is microcanonical on the energy surface, and recently studied
extensively in the Africa billiard. \cite{Li94}
However, in these references the average is taken
for a fixed wavefunction over a local region
around  $(\bbox{r} + \bbox{r}^\prime)/2$.

When an external magnetic field $B$ is present,
the electronic classical underlying dynamics undergoes a
transition from chaotic to integrable as the field gets stronger,
 regardless the shape of the billiard.
 In this paper, however,  we only discuss the case of weak fields for which
the motion is chaotic, and we are interested  in  the transition
 from orthogonal to unitary symmetry. While in the unitary case
 the wavefunctions
become complex, the arguments that lead to Eq. (\ref{15}) are still valid
 and the wavefunction correlator $C(k|\Delta\bbox{r}|)$ is unchanged.

The wavefunction correlation $C(k|\Delta\bbox{r}|)$ has been also derived
for weakly disordered systems using the supersymmetry technique
\cite{Prigodin95} in the unitary and orthogonal symmetries.
In addition Ref. \onlinecite{Prigodin95}  derives the joint
probability distribution for the intensity of an eigenfunction at
two different points.
We remark that the joint distribution of the wavefunction amplitude
at $\Lambda$ points $\bbox{r}_c$ is a special case of (\ref{11})
obtained for $\gamma_{c\lambda} \equiv \Psi_\lambda(\bbox{r}_c)$
(see the point contact case (\ref{5''}) except that the points $r_c$ can be
chosen anywhere within the dot and not only on the boundary).
We then obtain

\begin{eqnarray}
\label{17'}
 P(\Psi_\lambda(\bbox{r}_1), \Psi_\lambda(\bbox{r}_2), \ldots,
 \Psi_\lambda(\bbox{r}_\Lambda)) =
 (\det M)^{-\beta/2}
	     \, \exp \left[ - \frac{\beta}{2}
	    \sum\limits_{c c^\prime=1}^\Lambda \Psi^\ast_\lambda(\bbox{r}_c)
 \left(M^{-1}\right)_{c c^\prime}
\Psi_\lambda(\bbox{r}_{c^\prime})  \right]\;,
\end{eqnarray}
where $M_{cc^\prime} =  {\cal A}^{-1}J_0(k| \bbox{r}_c -
\bbox{r}_c^\prime|)$.
The distributions of  Ref. \onlinecite{Prigodin95} are then easily
obtained from (\ref{17'}) when $\Lambda=2$.\cite{Srednicki95}

\subsection{Total Width Distribution}

We calculate next the total width distribution $P(\Gamma)$ in a
given lead that supports $\Lambda$ channels and is characterized
by a correlation matrix $M$.
Although this quantity is not directly measurable in
 experiments with quantum dots, it appears very often in resonant scattering
by complex objects. \cite{Alt95,compnuc}
We remark that for a  dot  with reflection symmetry
$\Gamma^l=\Gamma^r \equiv \Gamma$
the conductance peak $g$ in (\ref{condpeak}) is proportional to
$\Gamma$. Such dots are, however, difficult to fabricate.

Using $\Gamma=\sum_c|\gamma_c |^2=\bbox{\gamma}^\dagger\bbox{\gamma}$,
the characteristic function of $P(\Gamma)$ is given by

\begin{eqnarray}
\label{18}
 \widetilde{P}(t) =
   \int^{\infty}_{-\infty}  d \Gamma \exp (it\Gamma) P(\Gamma)
 = \int^{\infty}_{-\infty} D \left[\bbox{\gamma} \right]
	   \exp (it\bbox{\gamma}^\dagger\bbox{\gamma})\,
	   P(\bbox{\gamma}) \;.
\end{eqnarray}
Since $P(\bbox{\gamma})$ is a Gaussian, we readily obtain
$\widetilde{P} (t)=\left[\det (I -2iMt/\beta)\right]^{-\beta/2}$.
The distribution itself is then given by an inverse Fourier transform

\begin{eqnarray}
\label{Pgamma}
 P(\Gamma) = \frac{1}{2 \pi} \int^{\infty}_{-\infty}
    dt \,\frac{e^{-i t \Gamma}}{\left[\det (I -
       2 i t M/\beta)\right]^{\beta/2}} \; .
\end{eqnarray}

The matrix $M$ is Hermitean and positive definite (since $\bbox{x}
^\dagger M \bbox{x}= \overline{| \bbox{x}\cdot \bbox{\gamma}|^2} >
0$ for any $\bbox{x} \neq 0$) and therefore its eigenvalues $w_c^2$
are all positive.
According to (\ref{Pgamma}), $P(\Gamma)$ depends only on $w_c^2$.
This is a consequence of the invariance of  $\Gamma$  under a orthogonal
(unitary) transformation of the $\Lambda$ partial width amplitudes.

We first discuss the simpler GUE case, for which the integrand has
poles $-i/w_c^2$ along the negative imaginary axis.
Taking a contour integration along the real line and  a half-circle
 that encloses all the poles in the lower half of the plane, we can calculate
(\ref{Pgamma}) by residues.
Assuming that all eigenvalues of $M$ are non-degenerate, the poles are
all simple and we find

\begin{eqnarray}
\label{PgamGUE}
 P_{GUE}(\Gamma) =
   \left(\prod_c\frac{1}{w_c^2}\right)
   \sum_{c=1}^\Lambda \left[\prod_{c^\prime \neq c}
   (\frac{1}{w_{c^\prime}^2} - \frac{1}{w_c^2})\right]^{-1}
   \mbox{e}^{-\Gamma/w_c^2} \;.
\end{eqnarray}
The distribution $P_{GUE}(\Gamma)$ given by (\ref{PgamGUE}) must be
positive, which can be directly verified by using the concavity of
the exponential function.

For two channels ($\Lambda=2$) which are in general non-equivalent
($M_{11} \neq M_{22}$) and correlated ($M_{12}\neq 0$), the
eigenvalues are given by $w_{1,2}^2 = (M_{11}+M_{22})/2 \pm
\sqrt{\left((M_{11}-M_{22})/2\right)^2 + | M_{12}|^2}$.
Then, Eq. (\ref{PgamGUE}) reduces to

\begin{eqnarray}
\label{21}
  P_{GUE}^{\Lambda=2}(\hat\Gamma) = \frac{2a_+}{\sqrt{a_-^2 + | f |^2}}
	\,\mbox{e}^{-2 a_+^2 \hat\Gamma/(1- |f|^2) }
	\sinh \left( \frac{2a_+ \sqrt{a_-^2 + |f|^2}}
	    {1 -|f |^2 }\hat\Gamma \right)\;,
\end{eqnarray}
where ${\hat\Gamma}=\Gamma/\overline{\Gamma}$ is the width in units
of its average value, $f=M_{12}/\sqrt{M_{11}M_{22}}$ measures the
degree of correlation between the two channels and $a_\pm =
1/2 \left(\sqrt{M_{11}/M_{22}} \pm \sqrt{M_{22}/M_{11}}\right)$
are dimensionless parameters such that for equivalent channels
$a_+=1$ and $a_-=0$.
In the latter case, we reproduce the result of Ref.
\onlinecite{Mucciolo95}.

For degenerate eigenvalues, we can calculate (\ref{Pgamma}) by using
the residue formula for higher order poles.
Alternatively we can slightly break the degeneracy of the eigenvalues
by $\eta$ and take the limit $\eta\rightarrow 0$.
For example, for two channels Eq. (\ref{PgamGUE}) gives $P(\Gamma)=
\left( \mbox{e}^{-\Gamma/w_1^2} - \mbox{e}^{-\Gamma/w_2^2} \right)/
(w_2^2-w_1^2)$.
By taking $w_2^2=w_1^2 + \eta$, in the limit $\eta \rightarrow 0$
we recover

\begin{eqnarray}
\label{23}
P(\Gamma) = \frac{\Gamma}{w^4} \, \mbox{e}^{-\Gamma/w^2} \;,
\end{eqnarray}
which is the  $\chi^2$ distribution in four degrees of
freedom. More generally, when all $\Lambda$ channels are uncorrelated
and equivalent  ($M=w^2 I$) we recover the well-known
 $\chi^2$ distribution in $2\Lambda$
degrees of freedom\cite{Porter65}

\begin{eqnarray}
\label{chi22L}
P_{GUE}^{(0)}(\Gamma) = \frac{1}{w^{2\Lambda} (\Lambda-1) !} 
 \Gamma^{\Lambda -1}\,
 \mbox{e}^{-\Gamma/w^2} \;.
\end{eqnarray}
 We have denoted this limiting distribution in (\ref{chi22L}) by $P_{GUE}^{(0)}$
as it will serve as our reference distribution against which to
compare the distributions in the general case of correlated and/or
inequivalent channels.

For the GOE case, the integral of Eq.(\ref{Pgamma}) is more
difficult to evaluate since the singularities of the integrand
along the negative imaginary axis $t=-i\tau$ are of the type
$(\tau -1/2w_c^2)^{-1/2}$.
In this case the semi-circle part of the contour (in the lower half
of the plane) is deformed to go up and then down along the negative
imaginary axis so as to exclude all the singularities.  When going around
a singularity of the above type the function changes sign.
 Therefore, after sorting the inverse eigenvalues of $M$ in
ascending order $w_1^{-2}< w_2^{-2} < \ldots$, we have

\begin{eqnarray}
\label{PgamGOE}
 P_{GOE}(\Gamma) = \frac{1}{\pi 2^{\Lambda/2}}
  \left(\prod_c  \frac{1}{w_c} \right)
 \sum_{m=1} \int_{1/2 w^2_{2m-1}}^{1/2 w^2_{2m}}
 d \tau \frac{\mbox{e}^{-\Gamma \tau}}
 {\sqrt{\prod_{r=1}^{2m-1}   (\tau - \frac{1}{2 w^2_r})
	\prod_{s=2m}^\Lambda (\frac{1}{2 w^2_{s}} - \tau)}} \;,
\end{eqnarray}
where for an odd number of channels $\Lambda$, we define
$1/2 w^2_{\Lambda+1} \rightarrow \infty $.
The integrand of each term on the  r.h.s. of (\ref{PgamGOE}) is
singular at the two endpoints of the integration interval, but
this singularity is integrable.
For the case of two channels that are in general non-equivalent but
correlated, Eq. (\ref{PgamGOE}) reduces to

\begin{eqnarray}
\label{25}
 P^{\Lambda=2}_{GOE}(\hat\Gamma) =
   \frac{a_+}{ \sqrt{1-|f|^2}}\,\mbox{e}^{-{a_+^2\hat\Gamma}/(1-|f|^2)}
   \; I_0\!\left(  \frac{a_+ \sqrt{a_-^2 + |f|^2}}
   {1 -|f |^2 }{\hat\Gamma} \right) \;,
\end{eqnarray}
where $f$ and $a_\pm$ are defined as before (see following Eq. (\ref{21}))
and $I_0$ is the Bessel function of order zero.
The case of equivalent channels is obtained in (\ref{25}) by
substituting $a_+=1$ and $a_-=0$.

 The reference distribution $P_{GOE}^{(0)}$, defined as before for the case
where  all $\Lambda$ channels are
equivalent and uncorrelated, is found directly from (\ref{Pgamma})
to be the $\chi^2$ distribution in $\Lambda$ degrees of
freedom\cite{Alhassid86,Jalabert92}

\begin{eqnarray}
\label{chi2L}
P_{GOE}^{(0)}(\Gamma) = \frac{1}{(2w^2) ^{\Lambda/2} (\Lambda/2 -1) !}
\Gamma^{\Lambda/2 -1}\, \mbox{e}^{-\Gamma/2w^2} \;.
\end{eqnarray}

The top panels (a and b) in Fig. \ref{fig1} show the width distributions
 for a two-channels lead in the GOE statistics.
The left panel is for equivalent channels ($M_{22}/M_{11}=1$) and
for various degrees of correlations $f=0.25, 0.5,0.75$, and $0.95$.
The right panel is for uncorrelated ($f=0$) but non-equivalent
channels: $M_{22}/M_{11}= 2, 3, 4$, and $5$.
The bottom panels (c and d) in Fig. \ref{fig1} are similar to
(a and b) except that they correspond to the GUE case.
We note that in all figures we display as $\Gamma$  the normalized
total width $\Gamma/\overline{\Gamma}$.

The correlation matrix in the point contact model is fully determined
by $k|\Delta \bbox{r}|$ and the number of channels $\Lambda$.
The left panels in Fig. \ref{fig2} show the GOE width distributions for
$k|\Delta \bbox{r}|=0.25,1, 4$ and for different number of channels
$\Lambda=2, 4$, and $6$.
The right panels of Fig. \ref{fig2} show similar results but for the GUE
statistics.
The deviation of the width distribution from the reference distribution
$P^{(0)}(\Gamma)$ which corresponds
to equivalent and uncorrelated channels
 (dashed lines in Fig. \ref{fig2})
becomes larger as the number of channels increases for a given
$k|\Delta \bbox{r}|$.

\subsection{Conductance Peaks Distribution}

To calculate the conductance distribution $P(g)$ in the general
case, we assume that the left and right leads are far enough
from each other and thus uncorrelated.  \cite{corrections}
The left and right leads are characterized by their own correlation
matrix  $M^{l}$ and $M^{r}$, respectively.
Under this assumption

\begin{eqnarray}
\label{Pg}
P(g) = \int\!d\Gamma^l \, d\Gamma^r \,
   \delta\!\left(g - \frac{\Gamma^l \Gamma^r}{\Gamma^l + \Gamma^r}\right)
   P(\Gamma^l)\,P(\Gamma^r) \;,
\end{eqnarray}
where $P(\Gamma)$ is given by (\ref{PgamGUE}) in the unitary case
and by (\ref{PgamGOE}) in the orthogonal case.

The distribution $P(g)$ can be evaluated by the following identity

\begin{eqnarray}
\label{29}
 \int_0^\infty d\Gamma_1\int_0^\infty d\Gamma_2\,
&\mbox{e}^{-\Gamma_1/\delta_1} \mbox{e}^{-\Gamma_2/\delta_2}  &
 \,\delta\!\left(g - \frac{\Gamma_1\Gamma_2}{\Gamma_1 + \Gamma_2}\right)
\nonumber \\
  = &  4g \,\mbox{e}^{-\left( {1\over \delta_1 } +
       {1\over\delta_2} \right) g} &
 \left[ K_0\left({2g\over \sqrt{\delta_1\delta_2}} \right) +
{1\over 2} \left( \sqrt{{\delta_2\over\delta_1}} +
\sqrt{{\delta_1\over\delta_2}}\right) K_1\left(
 {2g\over \sqrt{\delta_1\delta_2}} \right) \right] \;,
\end{eqnarray}
provided $\delta_1, \delta_2 > 0$. To obtain this identity we have used
the integral representation of the Bessel function
$K_\nu(z) = 1/2 (z/2)^\nu \int_0^\infty \! dt\,t^{-\nu-1} e^{-t-z^2/4t}$\,.

For the unitary case, Eqs. (\ref{PgamGUE}) and (\ref{29}) give

\begin{eqnarray}
\label{PgGUE}
 P_{GUE}&&(g) =
    16 g \left(\prod_c \frac{1}{v_c^2}\right)
	 \left(\prod_d \frac{1}{w_d^2}\right)
    \sum_{c,d}^{\Lambda,\Lambda^\prime}  \left[
    \prod_{c^\prime \neq c}(\frac{1}{v_{c^\prime}^2}-\frac{1}{v_c^2})
    \prod_{d^\prime \neq d}(\frac{1}{w_{d^\prime}^2}-\frac{1}{w_d^2})
	\right]^{-1}
  \nonumber \\ &&
    \times \,\mbox{e}^{-(\frac{1}{v_c^2} + \frac{1}{w_d^2})g}
     \left[ K_0\!\left(\frac{2g}{v_c w_d}\right)
     + \frac{1}{2}\left(\frac{ v_c }{ w_d}+\frac{ w_d}{ v_c} \right)
	K_1\!\left( \frac{2g}{v_c  w_d} \right) \right] \;,
\end{eqnarray}
where $v^2_c$ and $w^2_d$ are the eigenvalues of the left and right
lead correlation matrices $M^l$ and $M^r$, respectively.
The previous published results\cite{Jalabert92,Prigodin93} are
special cases of Eq.~(\ref{16}) for one channel leads with
$\overline{\Gamma}^l = \overline{\Gamma}^r$ ({\sl i.e.}  $v_1=w_1$),
while the distribution of Ref. \onlinecite{Mucciolo95} is obtained
for two (equivalent) channels leads whose matrices are related by
an overall asymmetry factor $M^r=a M^l$.

A similar calculation for the orthogonal limit gives

\begin{eqnarray}
\label{PgGOE}
 P_{GOE}&&(g) =
      \frac{4g}{\pi^2 2^\Lambda} \left(\prod_c \frac{1}{v_c}\right)
				 \left(\prod_d \frac{1}{w_d}\right)
      \sum_{m,m^\prime}
 \int_{1/2 v^2_{2m-1}       }^{1/2 v^2_{2m}       } d\tau
 \int_{1/2 w^2_{2m^\prime-1}}^{1/2 w^2_{2m^\prime}} d\tau^\prime
       \,\mbox{e}^{-(\tau - \tau^\prime) g}
 \nonumber \\
  & &
\times\frac{
     K_0 (2g \sqrt{\tau\tau^\prime})
       + \frac{1}{2}\left( \sqrt{\frac{\tau}{\tau^\prime}} +
			   \sqrt{\frac{\tau^\prime}{\tau}} \right)
     K_1 (2g \sqrt{\tau\tau^\prime})  }
  {\left[\prod_{r=1}^{2m-1}  \left(\tau - \frac{1}{2 v^2_r}\right)
	 \prod_{s=2m}^\Lambda\left(\frac{1}{2 v^2_s} - \tau\right)
	 \prod_{r^\prime=1}^{2m^\prime-1}
		   \left(\tau^\prime-\frac{1}{2 w^2_{r^\prime}}\right)
	 \prod_{s^\prime=2m^\prime}^\Lambda
		   \left(\frac{1}{2 w^2_{s^\prime}}-\tau^\prime\right)
  \right]^{1/2}}\;.
\end{eqnarray}

Fig. \ref{fig3} shows the GOE (left) and GUE (right) conductance
peak distribution (\ref{PgGOE}) and (\ref{PgGUE}), respectively,
for symmetric $\Lambda$-point leads with  $k|\Delta \bbox{r}|=0.25,
1, 4$ and for $\Lambda=2, 4$, and 6 (the same cases shown in Fig.
\ref{fig2}).
In analogy to $P(\Gamma)$, all figures depicting $P(g)$ display the
normalized conductance $g$ defined as $g/\overline{g}$.
By comparing Fig. \ref{fig3} with Fig. \ref{fig2} we conclude that, 
as $\Lambda$ increases, the conductance
distribution shows stronger deviation from its limiting case of
uncorrelated equivalent channels (dashed lines) than the width
distribution does.

Fig. \ref{fig4} shows the case of asymmetric leads for the asymmetry
factor  $a=1$ and 10, for four-point leads with $k|\Delta \bbox{r}|
= 1$ and for the orthogonal (a) and the unitary (b) limits.
$P(g)$ is not very sensitive to the leads asymmetry and a large
value of $a$ is needed to see significant variation from the
symmetric leads case.
In the limit $a \rightarrow \infty$, one can neglect the smaller
width in (\ref{condpeak}) and the conductance peak $g$ is proportional
to the partial width in the dominating lead.
In this limit $P(g)$ is reduced to $P(\Gamma)$ shown by the dashed
lines in Fig. \ref{fig4}).
The asymmetry effect becomes larger for an increasing number of channels.
 This effect can be noticed by comparing the GOE and
GUE cases, since for the same number of physical channels $\Lambda$
the GUE has a larger number of ``effective'' channels.

\section{Dynamical Model}

To test the RMT predictions for the statistical distributions,
we modeled a quantum dot by a system whose classical dynamics
is chaotic.
The model is the conformal billiard, \cite{Robnik83,Berry86}
whose shape is defined by the image of the unit circle  in
the complex $z$-plane under the conformal mapping
\begin{eqnarray}
\label{mapping}
w(z) = \frac{z+bz^2+c\mbox{e}^{i\delta}z^3}{\sqrt{1+2b^2+3c^2}} \;.
\end{eqnarray}
The parameters $b, c$ and $\delta$ control the billiard shape.
Eq. (\ref{mapping}) ensures that area ${\cal A}$ enclosed by $w(z)$
is normalized to $\pi$ and is independent of the shape.
We analyze the case $b=0.2$, $c=0.2$ and $\delta=\pi/2$, for which the
classical phase space is known to be chaotic. \cite{Bruus94}
We have verified that the corresponding spectrum exhibits  GOE-like
spectral fluctuations (we used 300 converged levels by diagonalizing a
matrix of order 1000).
This is demonstrated in Fig. \ref{fig5}, where the nearest-neighbors level
spacing distribution $P(s)$ and the $\Delta_3$ statistics,
which measures the spectral rigidity, are shown.

To investigate the effect of an external magnetic field, we
consider the same billiard threaded by an Aharonov-Bohm flux
line, \cite{Berry86,Bruus94} which does not affect the classical
dynamics.
The flux is parametrized by  $\Phi = \alpha \Phi_0$ where
$\Phi_0$ is the unit flux.
We use the same set of values for $b, c$, and $\delta$ to insure
classical chaotic motion, and choose $\alpha=1/4$ for maximal
time-reversal symmetry breaking.
The statistical tests shown in Fig.~\ref{fig5} confirm that this
choice of $\alpha$ corresponds to the unitary limit .
We  remark that  the $\Delta_3$-statistics is a better measure to
 distinguish between the GOE and GUE cases
than the level spacing distribution $P(s)$  (used in
 Ref. \onlinecite{Bruus94}.)

\subsection{Spatial Correlations}

The eigenfunction amplitude correlation $C(k |\Delta \bbox{r}|)=
\overline{\Psi^*(\bbox{r}) \Psi(\bbox{r}^\prime)}/\overline{|\Psi(
\bbox{r})|^2}$  was recently
investigated thoroughly for the conformal billiard. \cite{Li94}
The results agree fairly well with the theoretical prediction, namely
$C(k |\Delta \bbox{r}|)= J_0(k|\bbox{r}-\bbox{r}^\prime|)$, \cite{Berry77}
if one averages over the orientation of $\Delta \bbox{r}$.
This result is obtained based on semiclassical arguments and the
eigenfunctions studied in Ref.\onlinecite{Li94} were chosen
accordingly to be highly excited states
 (deep in the semiclassical region).

In order to apply this result to quantum dots, further considerations
are in order.
First, a typical semiconductor quantum dot in the submicrometer
range contains several hundred electrons, and it is therefore
not obvious that the eigenstates around the Fermi level are
necessarily semiclassical.
Second, scars associated with isolated periodic orbits
give corrections to  $C(k|\Delta \bbox{r}|)$ which depend on the
orientation of  $\Delta \bbox{r}$ and are of order
 $O(\hbar^{1/2})$.
 The fluctuations of the spatial correlation of the billiard eigenfunctions
were recently studied\cite{Srednicki96} and found also to be suppressed
by  $O(\hbar^{1/2})$.  These corrections are negligible if one averages
 over all orientations around a given point $\bbox{r}$, keeping the modulus
$|\Delta\bbox{r}|$ fixed, but this is difficult to implement experimentally.
 At a fixed orientation the fluctuations of the spatial correlations
seem to be rather small if  
$k|\Delta\bbox{r}| {\ \lower-1.2pt\vbox{\hbox{\rlap{$<$}
\lower5pt\vbox{\hbox{$\sim$}}}}\ } 3$
 so that (\ref{17}) is
a good approximation. For larger values of  $k|\Delta\bbox{r}|$, there could
 be significant fluctuations from (\ref{17}) but in this region the
 width and conductance distributions are closer to their limiting case of
independent channels and are not very sensitive to the exact
correlations.

Our results were obtained by using the billiard eigenfunctions with
Neumann boundary conditions where the normal derivative of the
wavefunctions vanishes on the boundaries.
We analyze eigenfunctions in the vicinity of the 100$^{th}$ excited level
which resembles the experimental situation.
By moving the points around
the circle we generate more statistics and average over orientations.
The results are shown in Fig.~\ref{fig6} where the correlations in the
model (solid line) compare well with the theoretical result (dashed
line) for both cases with and without magnetic flux.
The agreement is fair, particularly for $k|\Delta\bbox{r}| 
{\ \lower-1.2pt\vbox{\hbox{\rlap{$<$}\lower5pt\vbox{\hbox{$\sim$}}}}\ }5$.
For $k|\Delta\bbox{r}|\gg 1$, the deviations from the theoretical
value of $C(k|\Delta\bbox{r}|)$ are not important  since the
channels are weakly correlated and the distributions are very close
to those describing uncorrelated channels.
Thus, corrections to our analytical findings should not be large,
as is supported by the numerical evidence presented below.

 In our model studies we imposed Neumann boundary conditions
 around the entire billiard  and not just at the dot-lead boundary.
To mimic the experimental situation  we would have to impose mixed
boundary conditions, \cite{Bruus94} which makes the calculations
much more computationally  intensive. However, we now argue
that our simplified situation still provides reasonable results.
For extended leads,  the length of the dot-lead contact region
$D$ must satisfy $k D \gg 1$  in order to support
open transverse channels (in dots containing several hundred electrons).
Therefore, deviations from $C(k|\Delta\bbox{r}|)$ at the edge of
the dot-lead contact region (where our boundary conditions are
 unrealistic) are averaged out.
For point-like contacts, the physical picture is that the conductance
is probing the wavefunction in the vicinity of the constriction (the
region that couples the dot to the external lead).
 We then need  to know
the characteristic properties of the wavefunction inside the dot
where our model is quite satisfactory.

\subsection{Coupling to Leads and Distributions}

We first studied the point-like contacts model by describing the
lead as a sequence of $\Lambda$ equally spaced points on the boundary
of the billiard (in the $w$-plane).
According to Eq.~(\ref{17}) the correlation matrix $M$ is then
completely determined by $k |\Delta \bbox{r}| \approx k\delta\theta
|w^\prime(r=1, \theta)|$ (where $\delta \theta$ is the angle that
spans the arc between two neighboring points in the $z$-plane)
 and $\Lambda$.
In this model it is easy to generate strong correlations by choosing
the points close enough, unlike the (discretized) Anderson  model
\cite{Mucciolo95} where the channels are weakly correlated even
if the lead is composed of nearest neighboring points.
The eigenvalues $w_c^2$ are found by diagonalizing the matrix $M$.

In Figs. \ref{fig7} and \ref{fig8} we compare for the unitary and
orthogonal limits, respectively, the total width distribution
$P(\Gamma)$ obtained by solving the conformal billiard (histograms)
with the theoretical predictions (solid lines), for several values
of $k |\Delta \bbox{r}|=0.5,1,2$ and $\Lambda=2,4,6$.
The distributions $P^{(0)}(\Gamma)$ for equivalent and uncorrelated channels
are indicated by
the dashed lines, and are just the $\chi^2$ distributions in $\Lambda$
($2\Lambda$) degrees of freedom for the GOE (GUE).
The agreement between the model and the analytic RMT predictions confirms
the validity of the statistical model for a chaotic dot.
We observe from Figs. \ref{fig7} and \ref{fig8} that for the larger
values of $k |\Delta \bbox{r}|$, the distributions get closer to those
for uncorrelated channels.
This is consistent with the decrease in spatial correlations (see
Fig. \ref{fig7}).
Another interesting observation is that, for a constant $k |\Delta
\bbox{r}|$ ({\sl i.e.} fixed correlations), the deviation from the
limiting case of independent channels becomes larger with an increasing
number of channels.

Figs. \ref{fig9} and \ref{fig10} show a comparison between the
theoretical conductance peaks distributions for symmetric leads,
as given by Eqs.~(\ref{PgGUE}) and (\ref{PgGOE}) for the unitary
and orthogonal cases, respectively, and those calculated for the
Africa billiard with symmetric $\Lambda$-point leads ($\Lambda=2,4$,
and 6) and for different values of $k|\Delta \bbox{r}|$.
The dashed lines are again the limiting case of uncorrelated and
equivalent leads.
Observations that are similar to the ones made above for the
 width distributions, can be made with respect to the conductance peaks
 distributions.
Comparing the width and conductance peaks distributions, we note that
the conductance distribution shows stronger deviation from its limit
for uncorrelated equivalent channels than does the width distribution.

We also studied extended leads by taking the contact region of the lead
and the dot to have a finite length $D  \approx |w^\prime| \Delta \theta$
 on the dot's boundary (in the $w$-plane) where $w^\prime$ is evaluated
 at the corresponding angle where the lead is located.
In this case the channels are defined by the allowed quantized transverse
momenta $\kappa_c=\pi n_c/D$ with  $n_c=1, 2, \ldots, \Lambda$, where
$\Lambda=\mbox{int}[kD/\pi]$.
To calculate the partial amplitude for the conformal billiard, the
integral in Eq.~(\ref{gamdefR}) (defined in the $w$-plane) is mapped
into an integral along an arc in the $z$-plane which is spanned by
an angle $\Delta \theta$

\begin{eqnarray}
\label{27}
 \gamma_{c\lambda} = \sqrt{\frac{\hbar^2}{2m} }
     \int_{\Delta \theta} d\theta\,|w^\prime(r=1,\theta)| \,\Phi_c^*(\theta)
     \Psi_\lambda (r=1,\theta)  \;,
\end{eqnarray}
where $\Phi_c(\theta)= \sqrt{2/D} \sin (\kappa_c |w^\prime| \theta)$
are the transverse channel wavefunction and for simplicity we have set
$k_cP_c=1$.
The resonance eigenfunction $\Psi_\lambda$ is given in terms of
its expansion in $\mbox{e}^{i m \theta}$  (with $m= 0, \pm 1, \pm 2,
\ldots)$

\begin{eqnarray}
\label{27'}
  \Psi_\lambda(r=1,\theta) =
    {\cal N}_\lambda \sum_j \frac{c_j^\lambda}
     {\sqrt{\pi(\gamma_j^2 - |\ell_j-\alpha|^2) }}
    \,\mbox{e}^{i\ell_j \theta} \;,
\end{eqnarray}
where ${\cal N}_\lambda$ is a normalization constant, $\gamma_j$ are
the zeros of $J^\prime_{\mid\ell_j-\alpha \mid}$ and $c_j$ are expansion
coefficients as in Ref. \onlinecite{Bruus94}.

To guarantee that the correlation matrix $M$ in (\ref{8})  is the same for
eigenfunctions of the billiard which belong to different energies, we
choose $D$ such that
 $kD=constant$ and scale the partial amplitude
(\ref{gamdefR}) by $k^{1/2}$. The resulting matrix is

\begin{eqnarray}
\label{27''}
kM_{cc^\prime} = \frac{\hbar^2}{2m} {2 \over kD}
     \int_{\Delta \theta} d\theta\,   \int_{\Delta \theta}
d\theta^\prime\, & &|w^\prime(r=1,\theta)| \, |w^\prime(r=1,\theta^\prime)| \,
\nonumber \\
& & \times \sin \left({\pi n_c \over kD} |w^\prime| \theta\right)
     J_0 \left(|w^\prime| |\theta-\theta^\prime| \right)
\sin \left({\pi n_{c^\prime} \over kD} |w^\prime| \theta^\prime \right) \;.
\end{eqnarray}

This scaling is desirable in order to be consistent with the
theoretical approach presented above, but experimentally it is
 very hard to accomplish.
Fortunately, this scaling of $D$ is insignificant  for present experiments
\cite{Chang96,Folk96} that deal with dots containing several hundred
 electrons ${\cal N}$.
Indeed, from the Weyl formula we have
$k_F \propto {\cal N}^{1/2}$ so that  $\delta k_F / k_F =
\delta {\cal N} / 2{\cal N} \ll 1$. The latter inequality is obtained when
we estimate $\delta {\cal N}$ to be
the number of observed Coulomb blockade peaks
(since each Coulomb blockade peak corresponds
to the addition of one electron into the dot). The relative variation
of $k_F$ is thus small and can be neglected.

We find that the channels in the extended leads model are weakly correlated
and that the average partial widths in the various channels exhibit a
moderate variation.
In such a case the total width distribution is not very different from
the case of uncorrelated equivalent channels.
Our model calculations for extended leads are shown
 in Fig. \ref{fig11}  and are in agreement with the RMT predictions
for uncorrelated channels (dashed lines).
An interesting effect is that with an increasing number of channels
even small deviations in $P(\Gamma)$ give rise to relatively large
deviations  in $P(g)$, as can be seen in Fig.  \ref{fig11}d.

\section{Connection to experiments and conclusions}

We have discussed  both the cases of  orthogonal and unitary symmetries.
To relate to actual experimental situations, it is important to estimate
the minimal  strength of the magnetic field $B_c$ which ensures complete
time-reversal symmetry breaking.
 For a ballistic electron \cite{Berry86,Jalabert92,Bruus94,Frahm95}
 $B_c {\cal A} \propto \sqrt { \tau_{cr} / \tau}
 \Phi_0$, where $\tau_{cr}$ and $\tau$ are, respectively, the
time it takes the electron to cross the dot and the Heisenberg time
 $\tau=h / \Delta$. For an electron at the Fermi energy  $B_c {\cal A}
 \propto {\cal N}^{-1/4} \Phi_0$, where ${\cal N}$ is the number
 of electrons in the dot.
The proportionality factor is non-universal and
depends on the exact geometry of the dot.  In a semiclassical
analysis\cite{Bohigas95,Pluhar95} it can be expressed in terms of
classical quantities.
For the dots used in some recent experiments,
\cite{Chang96,Folk96} $B_c$ is of order of a few mT.
Such small values of $B_c$ do not alter significantly  the
 classical dynamics of the electron,\cite{Obs1} and  our assumption that
the  correlation $C(k|\Delta \bbox{r}|)$ is unchanged is justified.
 Nevertheless,  these small  variations in the magnetic field
have appreciable quantum mechanical effects, {\sl i.e.} the crossover
from orthogonal to unitary symmetry.

In conclusion, we have derived closed  expressions for the width and
conductance
peak distributions in quantum dots in the Coulomb blockade regime.
The main assumption is that the electron's dynamics is chaotic for
 a ballistic dot  or weakly diffusive for a disordered dot.
For given correlation matrices that characterize the left and right
leads, these distributions are universal and distinct for conserved
and broken time-reversal symmetry.
While recent experiments have measured the conductance distributions in
symmetric one-channel leads, it would be interesting to measure and
compare with theory  the conductance distributions in
more general situations of dots with multi-channel leads.


\acknowledgements

This work was supported in part by the U.S.  Department of Energy Grant
DE-FG02-91ER40608.
C.H.L. acknowledges financial support by the Conselho
Nacional de Pesquisas (CNPq -- Brazil). We thank
H.~U.~Baranger and A.D. Stone for discussions and H. Bruus
for the use of his billiard computer program.

%

\begin{figure}[p]

\vspace{5 mm}

\caption{
  Total width distributions $P(\Gamma)$ for a two-channel lead.
  Panels (a), (b) correspond to the orthogonal symmetry
 and (c), (d) to the unitary symmetry.
  $P(\Gamma)$ for equivalent but correlated channels
   with $f=0.25,0.5,0.75$ and $0.95$ are shown in (a) and (c), while
 uncorrelated but non-equivalent channels $M_{22}/M_{11}=2,3,4,5$
 are shown in (b) and (d).
  Descending values of $f$ (or $M_{22}/M_{11}$) correspond to
  distributions that extend to larger values of $\Gamma$. }
\label{fig1}

\vspace{5 mm}

\caption{
 Total width distributions for a $\Lambda$-point lead and
 $k |\Delta \protect\bbox{r}|= 0.25, 1$ and $4$ in a quantum dot with orthogonal
 symmetry (GOE) and with unitary symmetry (GUE).
 (a)  GOE; $\Lambda=2$; (c) GOE; $\Lambda=4$; (e) GOE; $\Lambda=6$;
 (b)  GUE; $\Lambda=2$; (d) GUE; $\Lambda=4$; (f) GUE; $\Lambda=6$.
 The dashed lines correspond to uncorrelated and equivalent channels.
 Increasing values of $k |\Delta \protect\bbox{r}|$ correspond to curves
 which approach the case of uncorrelated channels.}
\label{fig2}

\vspace{5 mm}

\caption{
Same as  Fig. \protect\ref{fig2} but for the conductance peak distributions
 $P(g)$ in dots with symmetric $\Lambda$-point
 leads. }
\label{fig3}

\vspace{5 mm}

\caption{
 Conductance peak distributions $P(g)$ for asymmetric four-point leads
 with $k|\Delta \protect\bbox{r}| = 1$ and an asymmetry factor of $a=1$
 and $a=10$.
 (a) GOE; (b) GUE.
 The dashed curves describe the limit $a \rightarrow \infty$ where
 $P(g)$ reduces to $P(\Gamma)$.}
\label{fig4}

\vspace{5 mm}

\caption{
  The nearest neighbors level spacing distribution $P(s)$ and the
  $\Delta_3$-statistics for the conformal billiard with $b=0.2,
  c=0.2$ and $\delta=\pi/2$.
  We consider the states between the 50th and the 350th.
  Left: no magnetic flux ( $\alpha=0$).
  Right: with magnetic flux of $\alpha=1/4$.}
\label{fig5}

\vspace{5 mm}

\caption{
 The spatial wavefunction correlation $C(k|\Delta \protect\bbox{r}|)$
 calculated for the conformal billiard (squares) compared with the
 theoretical prediction $J_0 (k|\Delta \protect\bbox{r}|)$ (solid line).
 Panel (a) displays the case where $\alpha=0$ and (b) corresponds to
 $\alpha=1/4$.}
\label{fig6}

\vspace{5 mm}

\caption {
  Total width distributions $P(\Gamma)$ for the unitary case for
  several values of  $k|\Delta \protect\bbox{r}| = 0.5,1,2$ and
  for various number of channels $\Lambda=2,4,6$. The solid lines
  are the theoretical distributions
   (\protect\ref{PgamGUE}), while the dashed lines correspond to
  uncorrelated and equivalent channels. The histograms are the
  results from the Africa billiard ($b=0.2, c=0.2, \delta=\pi/2$)
  where a magnetic flux line ($\alpha=1/4$) breaks the
  time-reversal symmetry. }
\label{fig7}

\vspace{5 mm}

\caption{
   Total width distributions $P(\Gamma)$ for conserved time-reversal
   symmetry (GOE).
   Conventions are as in Fig. \protect\ref{fig7}, with solid lines
   describing the orthogonal prediction  (\protect\ref{PgamGOE}).}
\label{fig8}

\vspace{5 mm}

\caption{
Conductance peak distributions $P(g)$ for the unitary symmetry.
The histograms display the results obtained from the Africa
billiard for symmetric $\Lambda$-point leads and different
values of $k |\Delta \protect\bbox{r}|$.
The solid lines are the RMT prediction (\protect\ref{PgGUE})
 and the dashed lines correspond to
uncorrelated and equivalent channels. The cases presented
are the same as in Fig. \protect\ref{fig7}. }
\label{fig9}

\vspace{5 mm}

\caption{
Same as Fig. \protect\ref{fig9} but for the orthogonal symmetry where
the theoretical distribution is given by Eq. (\protect\ref{PgGOE}).
}
\label{fig10}

\vspace{5 mm}

\caption{
Comparison of conformal billiard results for extended leads
with $\protect\mbox{int}[kD/\pi] = 6$ (histograms) and theoretical
predictions.
The panels represent total width distribution $P(\Gamma)$
for orthogonal (a) and unitary (b) limits and conductance
distribution $P(g)$ for orthogonal (c) and unitary (d)
limits.
The dashed lines correspond to uncorrelated equivalent
channels.}
\label{fig11}

\end{figure}


\begin{thebibliography}{99}
\bibitem{corrections}
    This approximation is equivalent to $kL \gg 1$, where $L$
    is the distance between the leads (approximately the size of the dot).
    It is hard to estimate the corrections of order $(kL)^{-1}$ to
    our results, and one would naively expect that they might impose
    limitations on our analytical findings.
    Fortunately the numerical results presented in the next section
    show that such corrections are indeed very small.
\bibitem{Obs1}
    Here we address the experimental situation where a uniform magnetic 
field is applied through the dot.  In the dynamical model analyzed in this 
paper the time-reversal symmetry is
    broken by a Bohm-Aharonov flux line which does not
    change at all the classical trajectories.
\end{thebibliography}
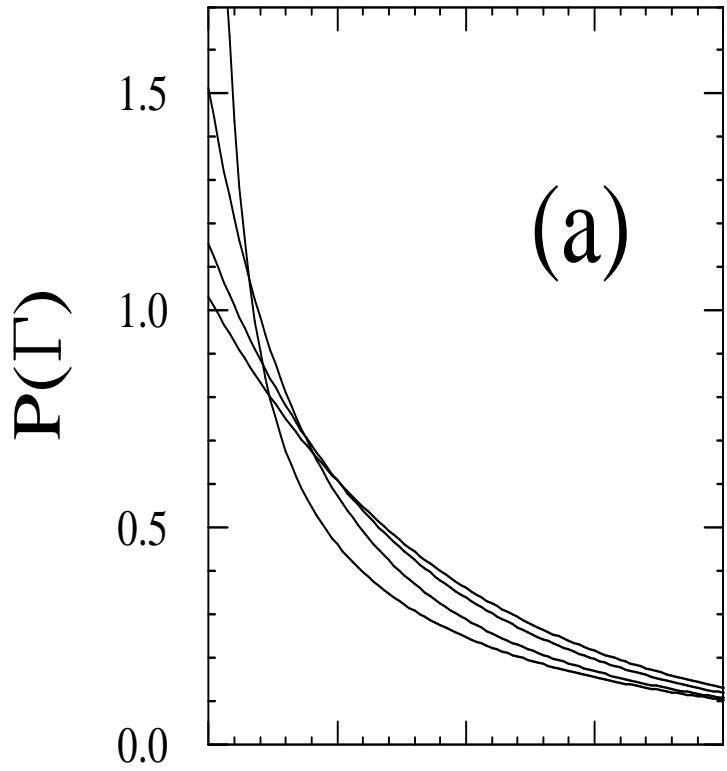
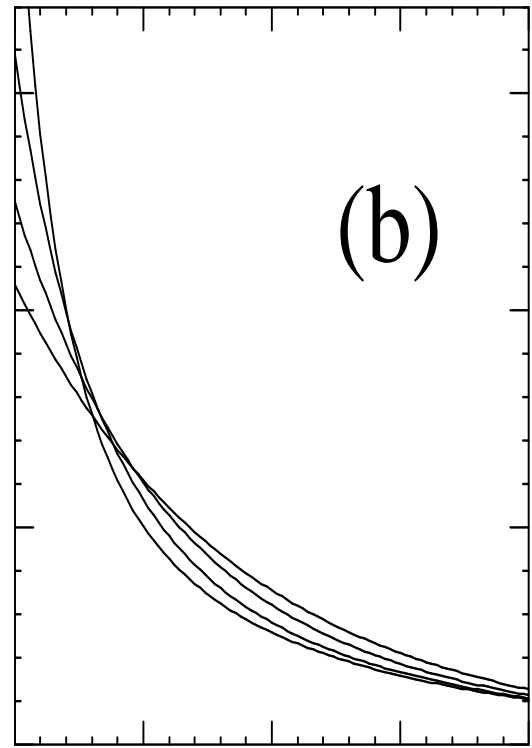
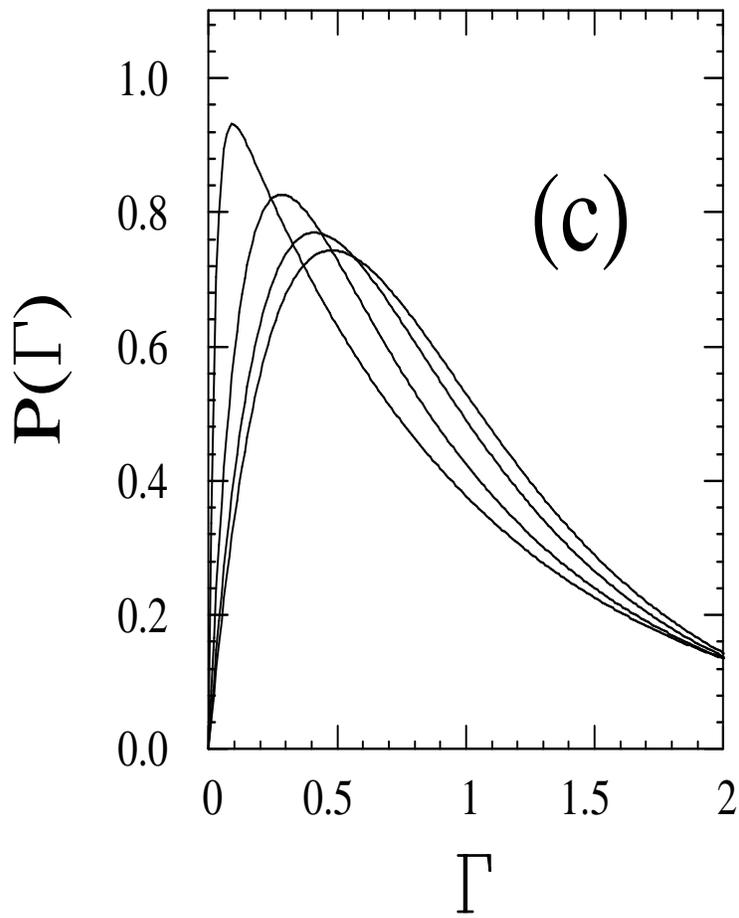
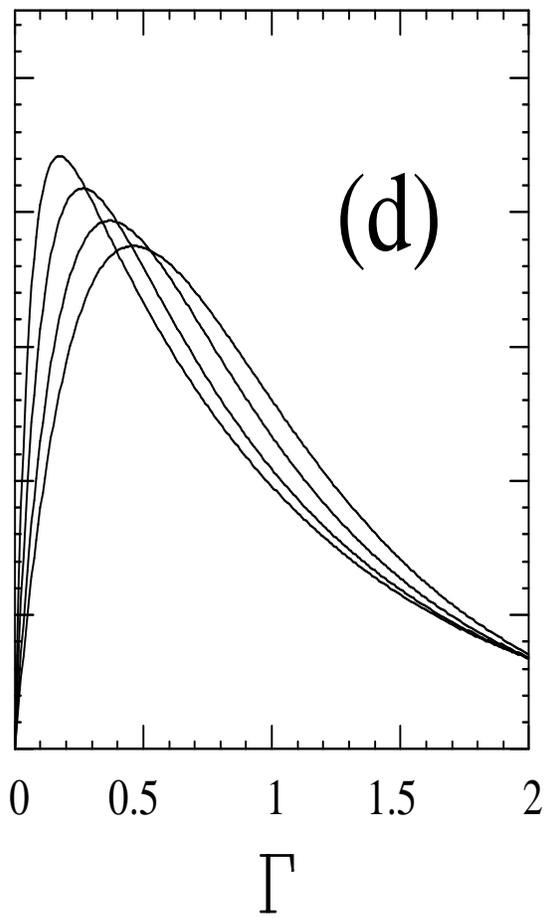
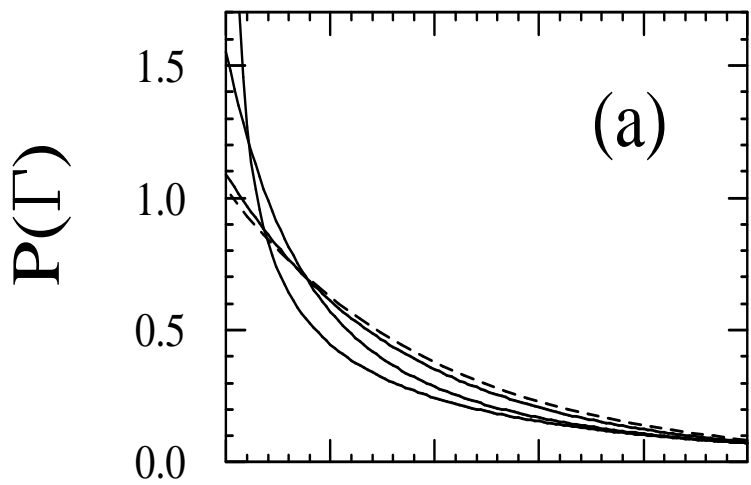
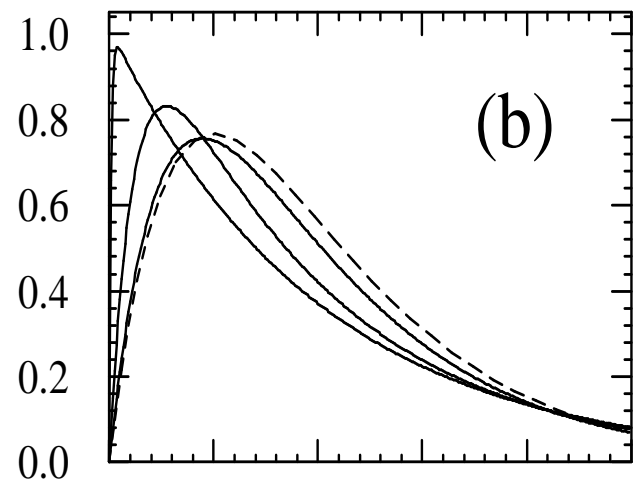
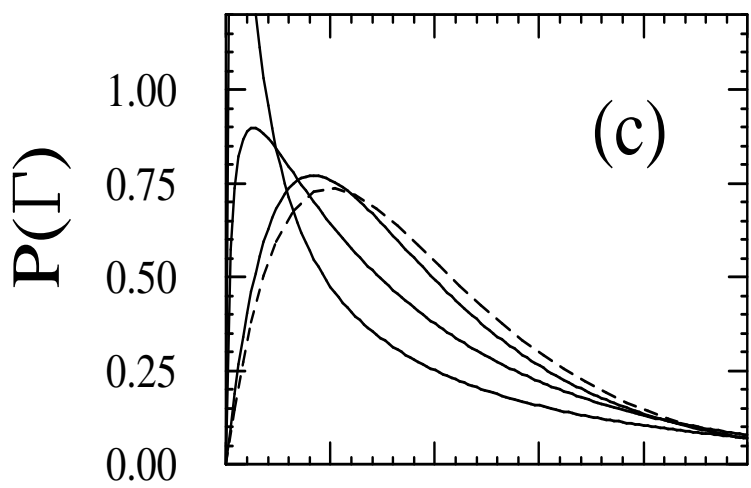
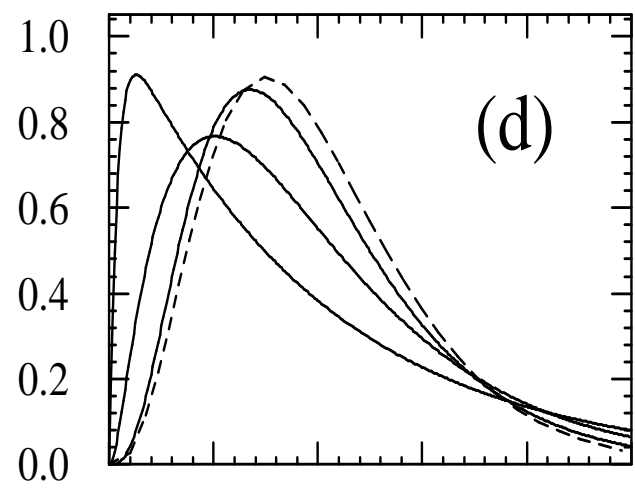
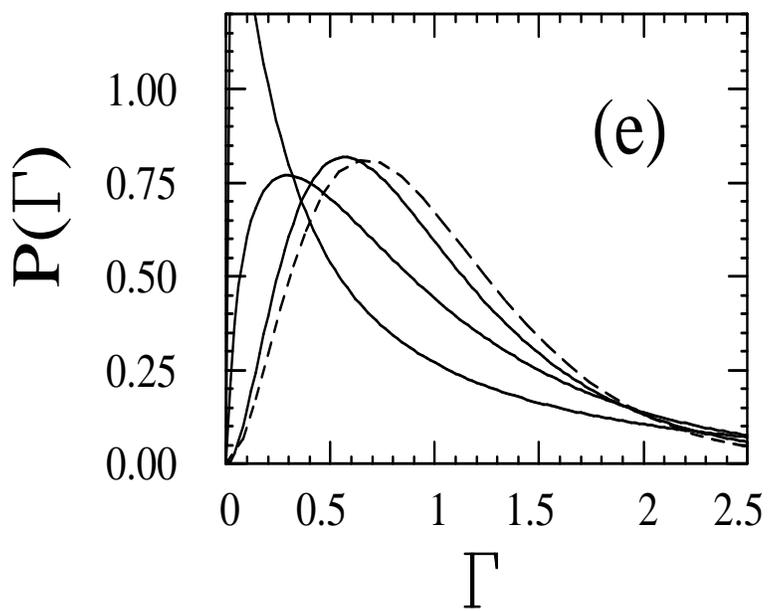
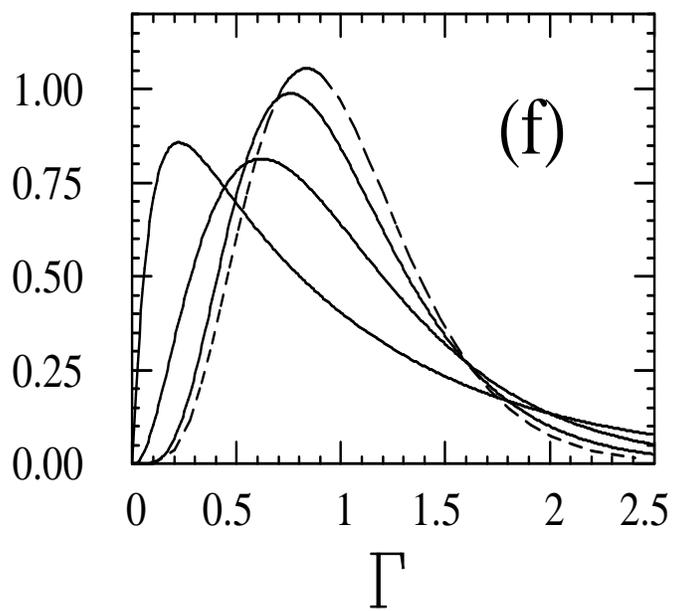
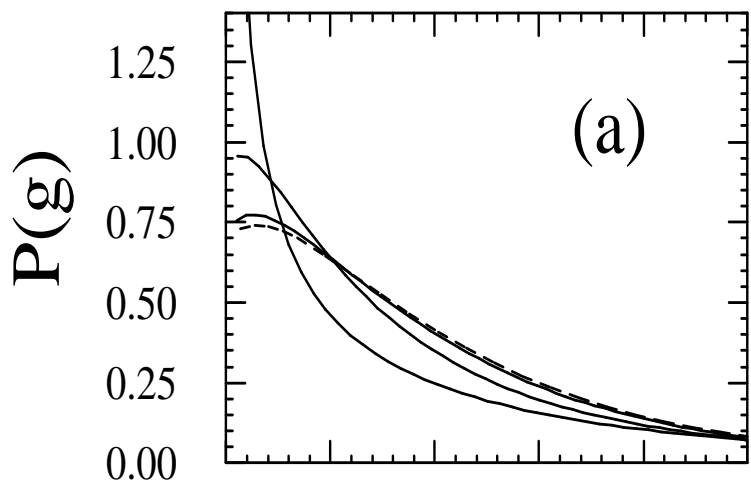
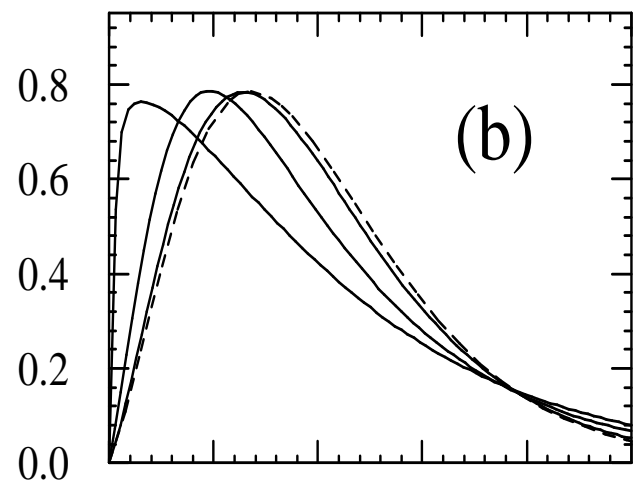
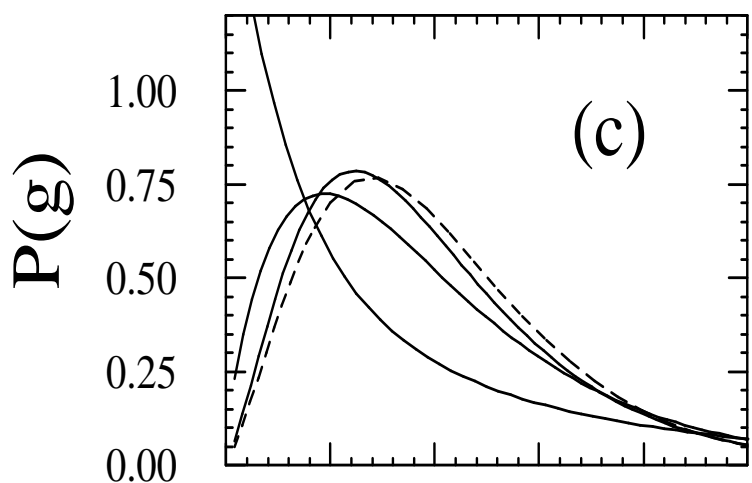
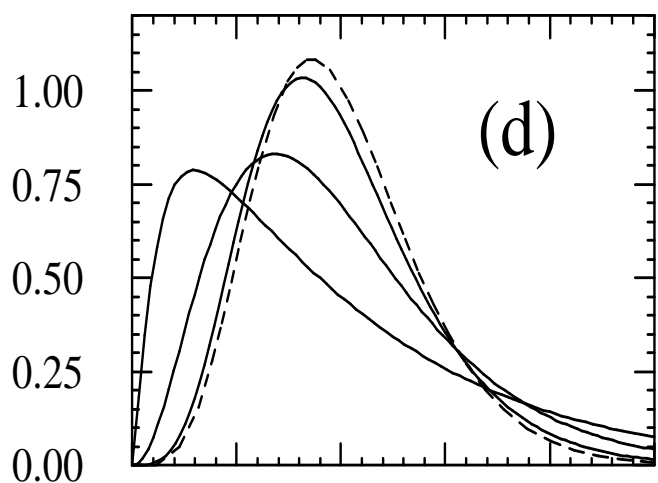
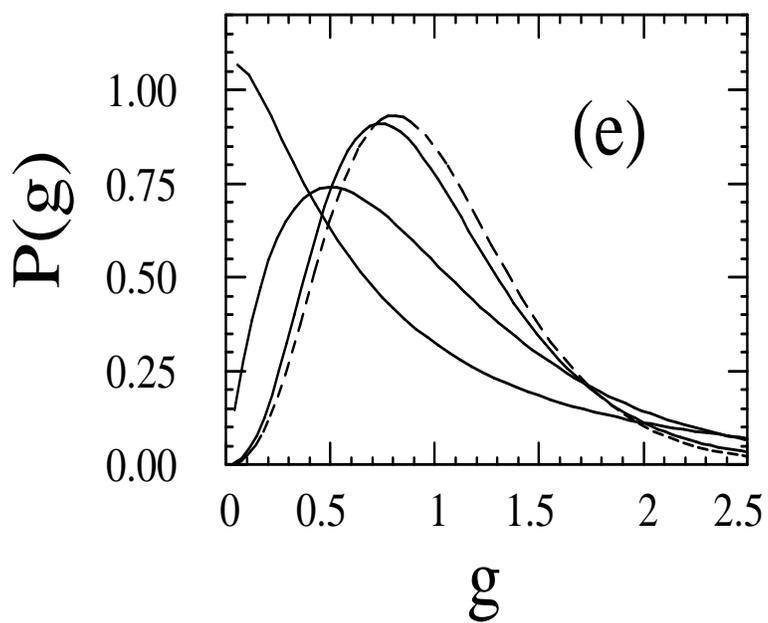
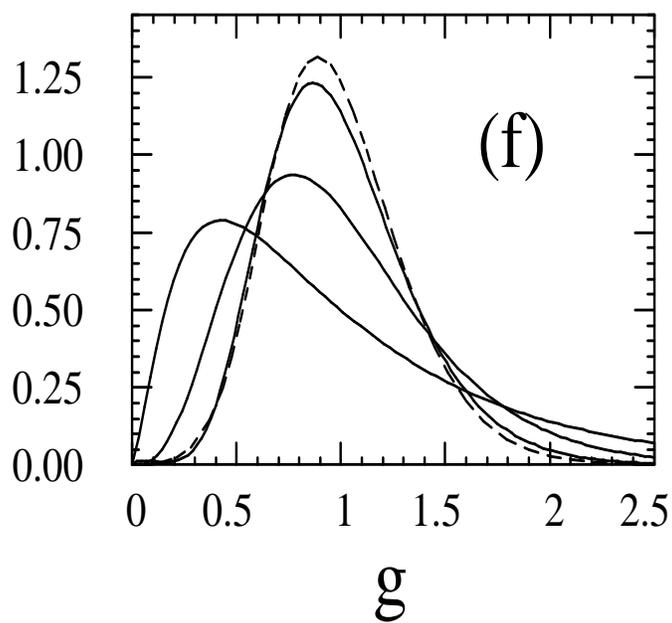
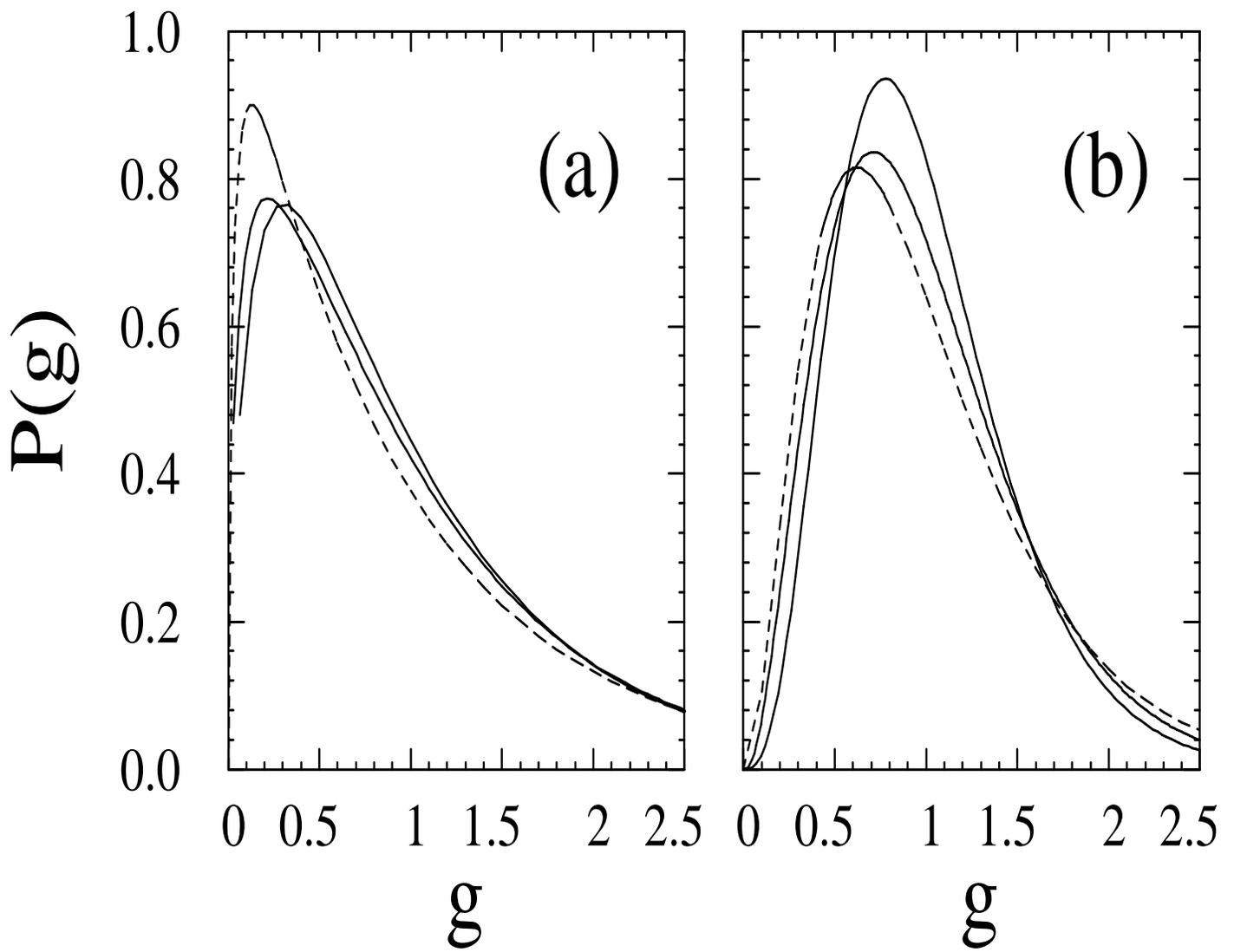
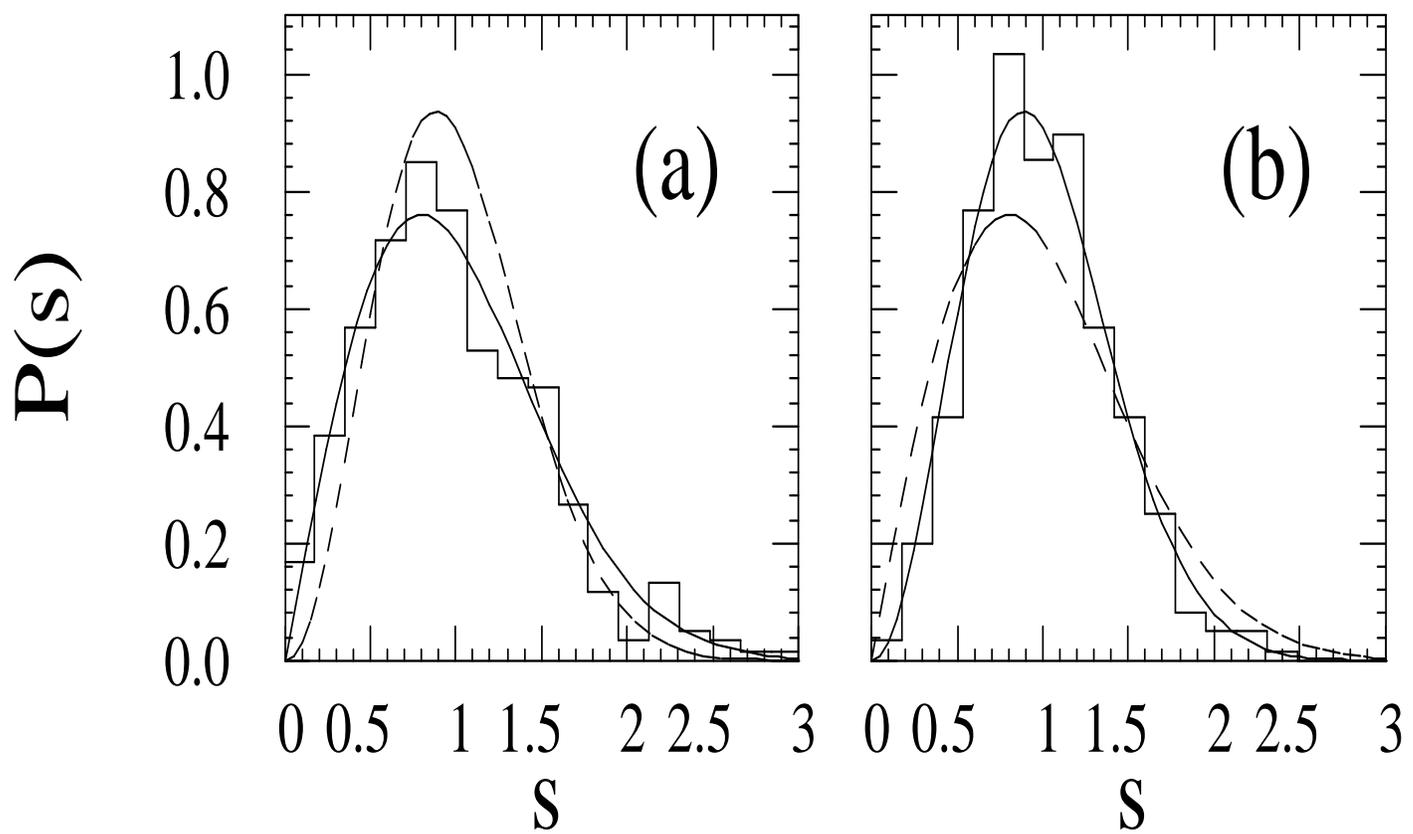
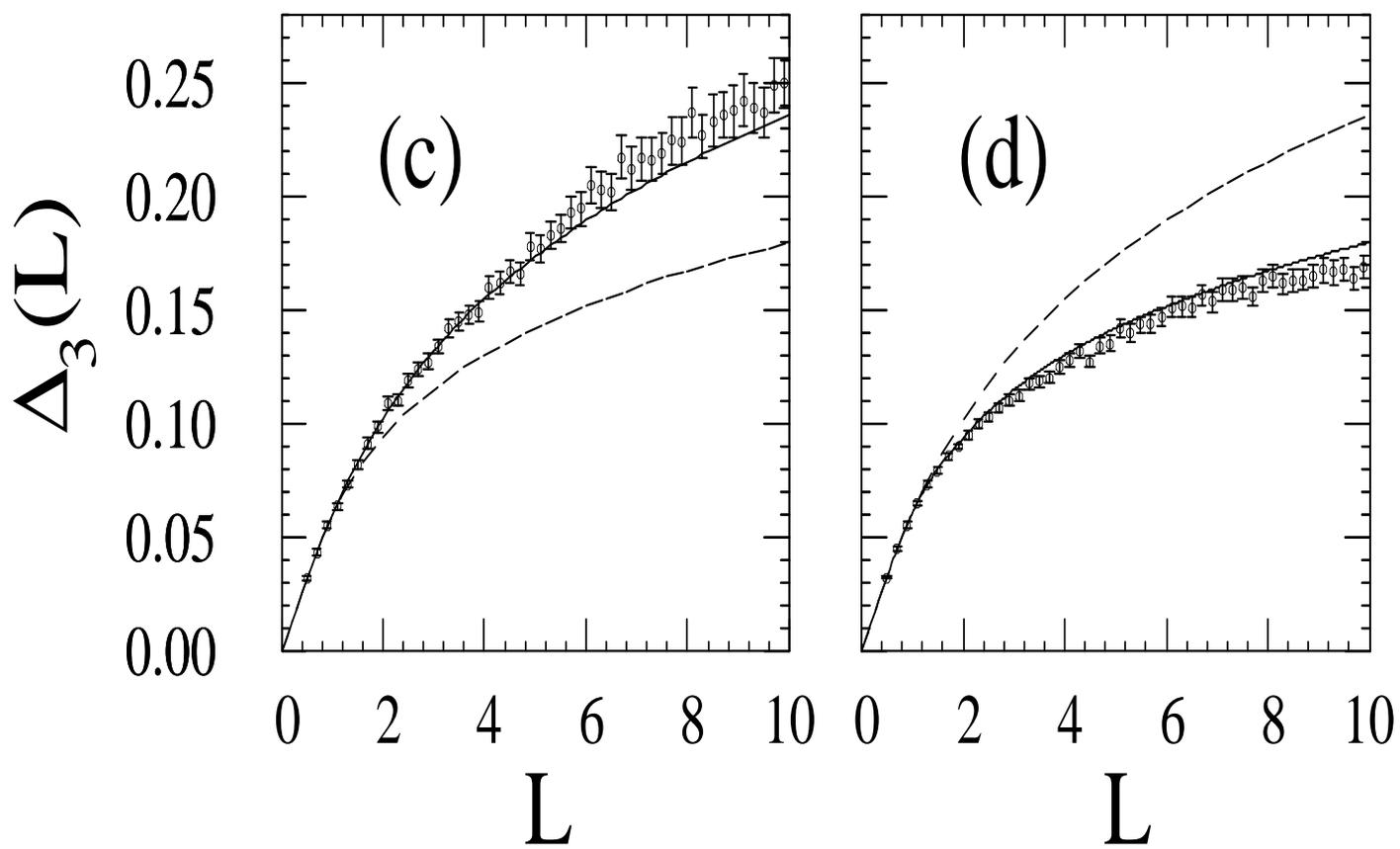
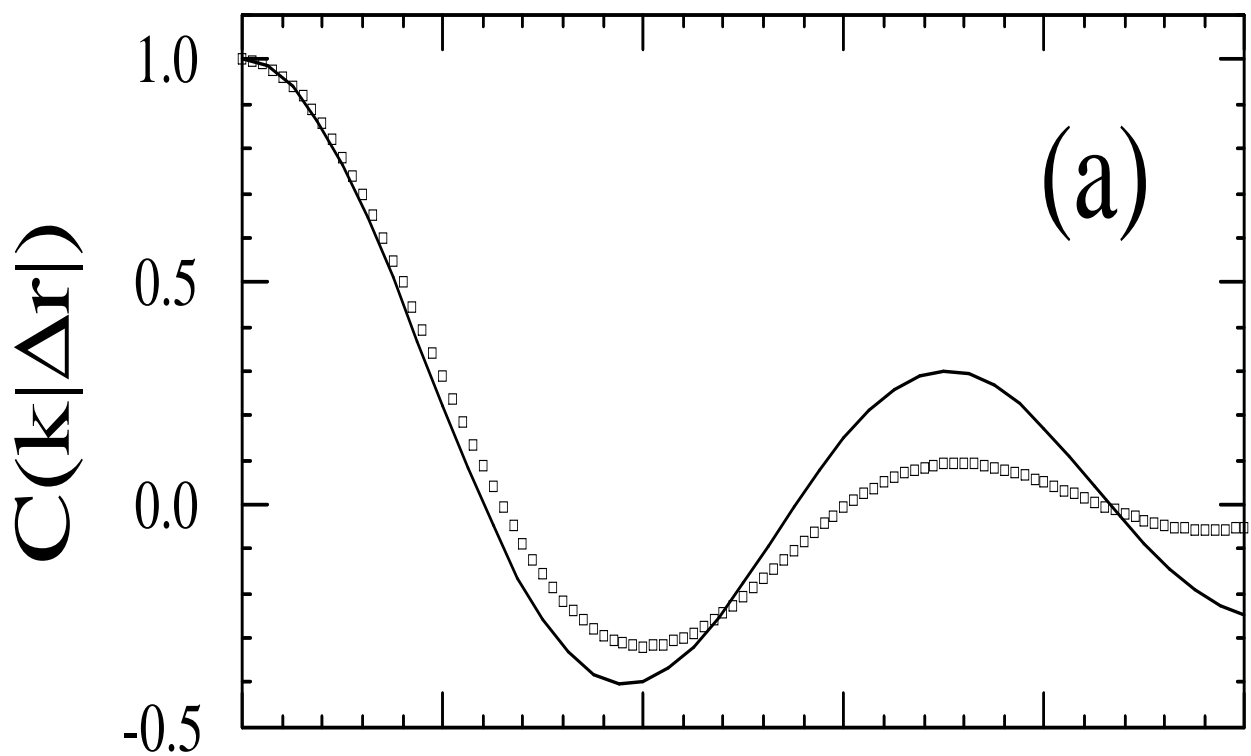
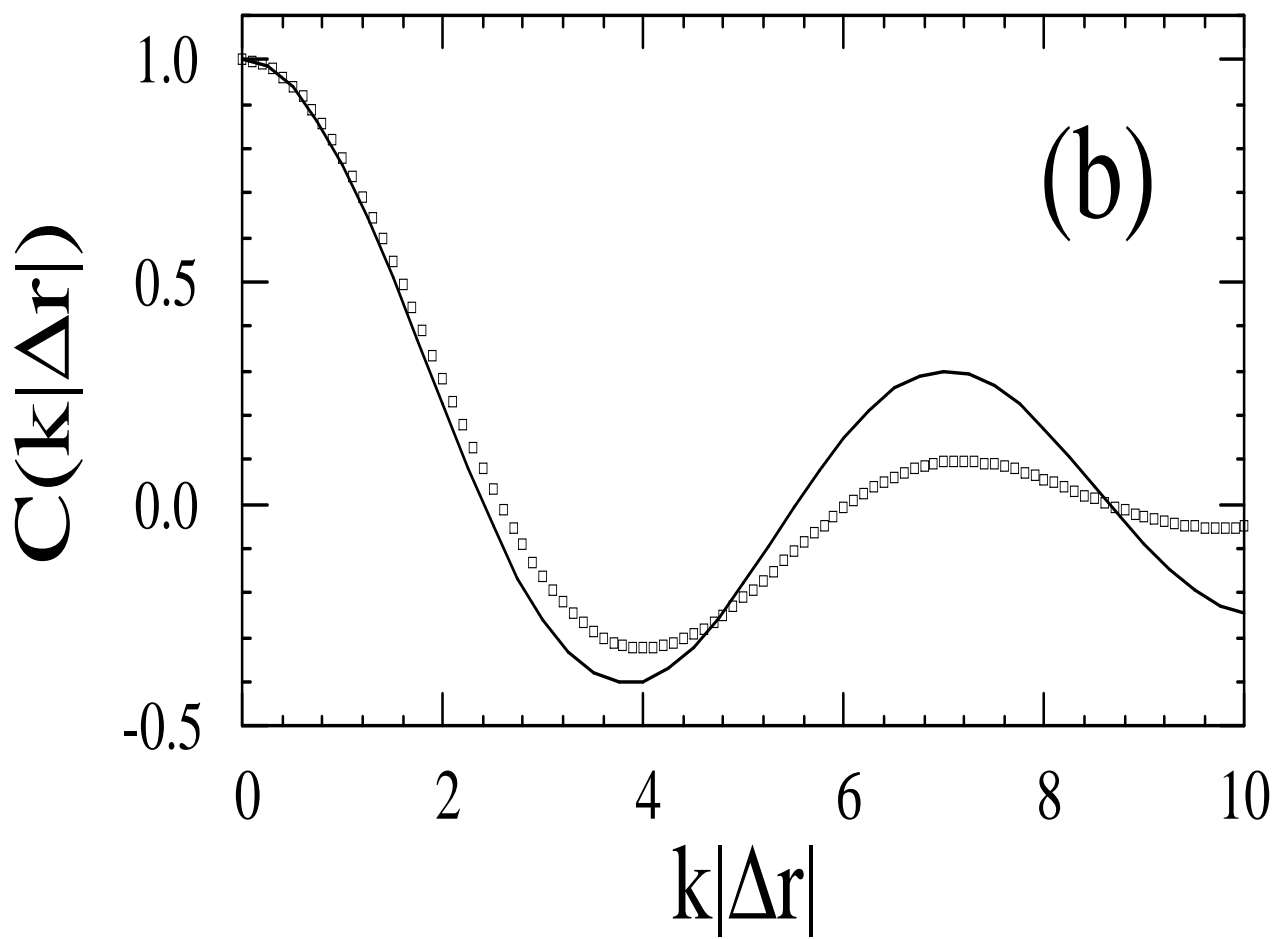
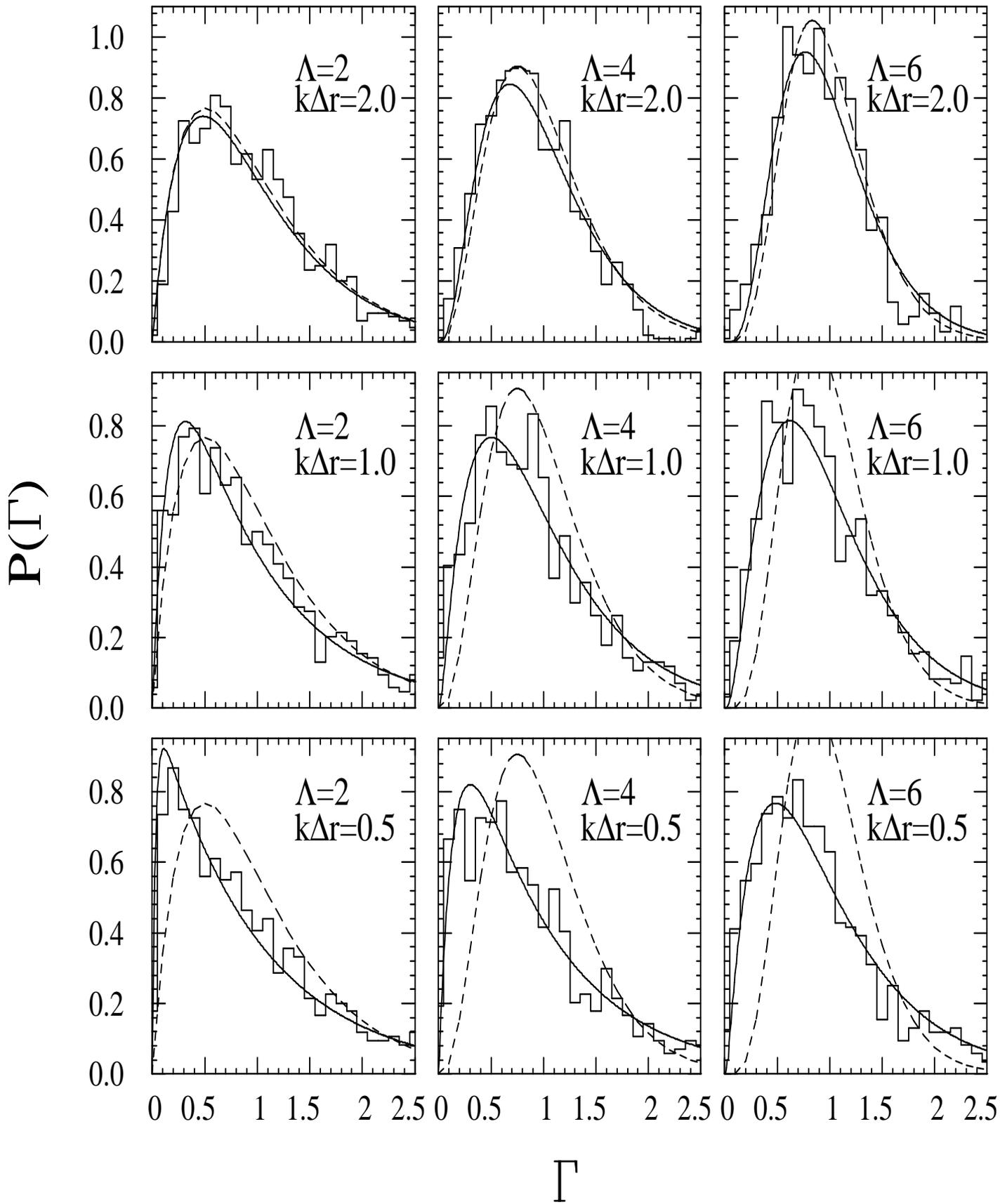
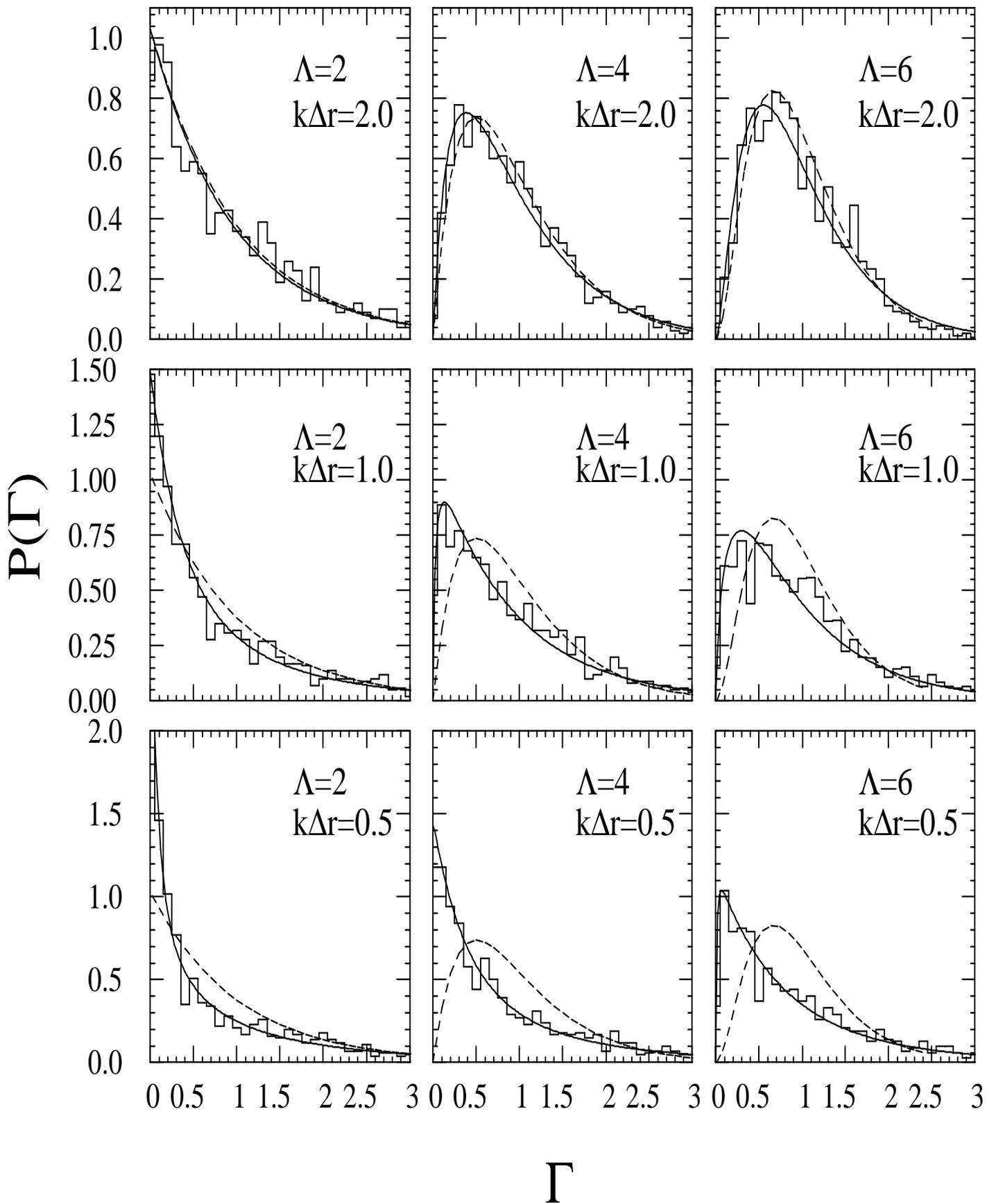
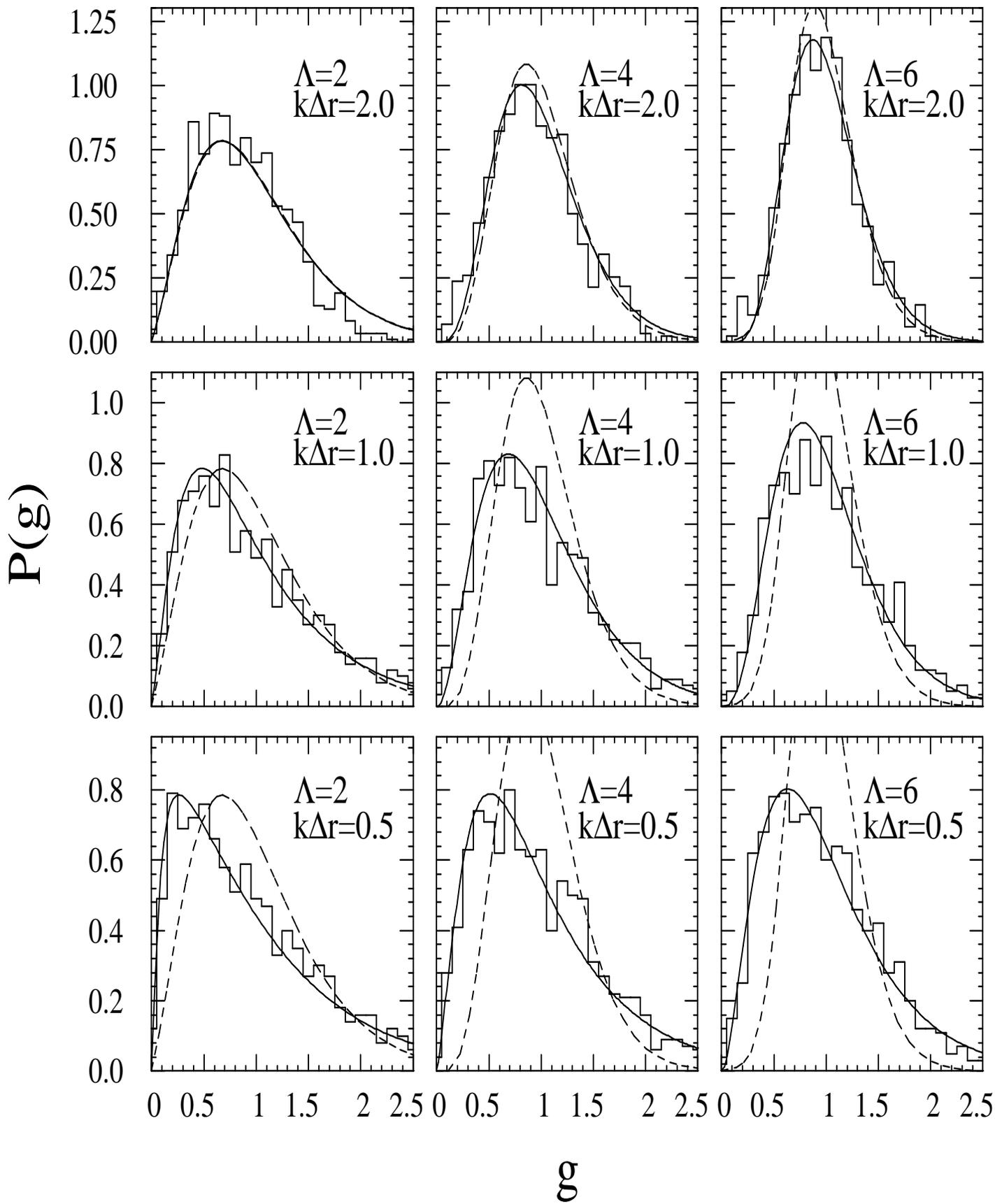
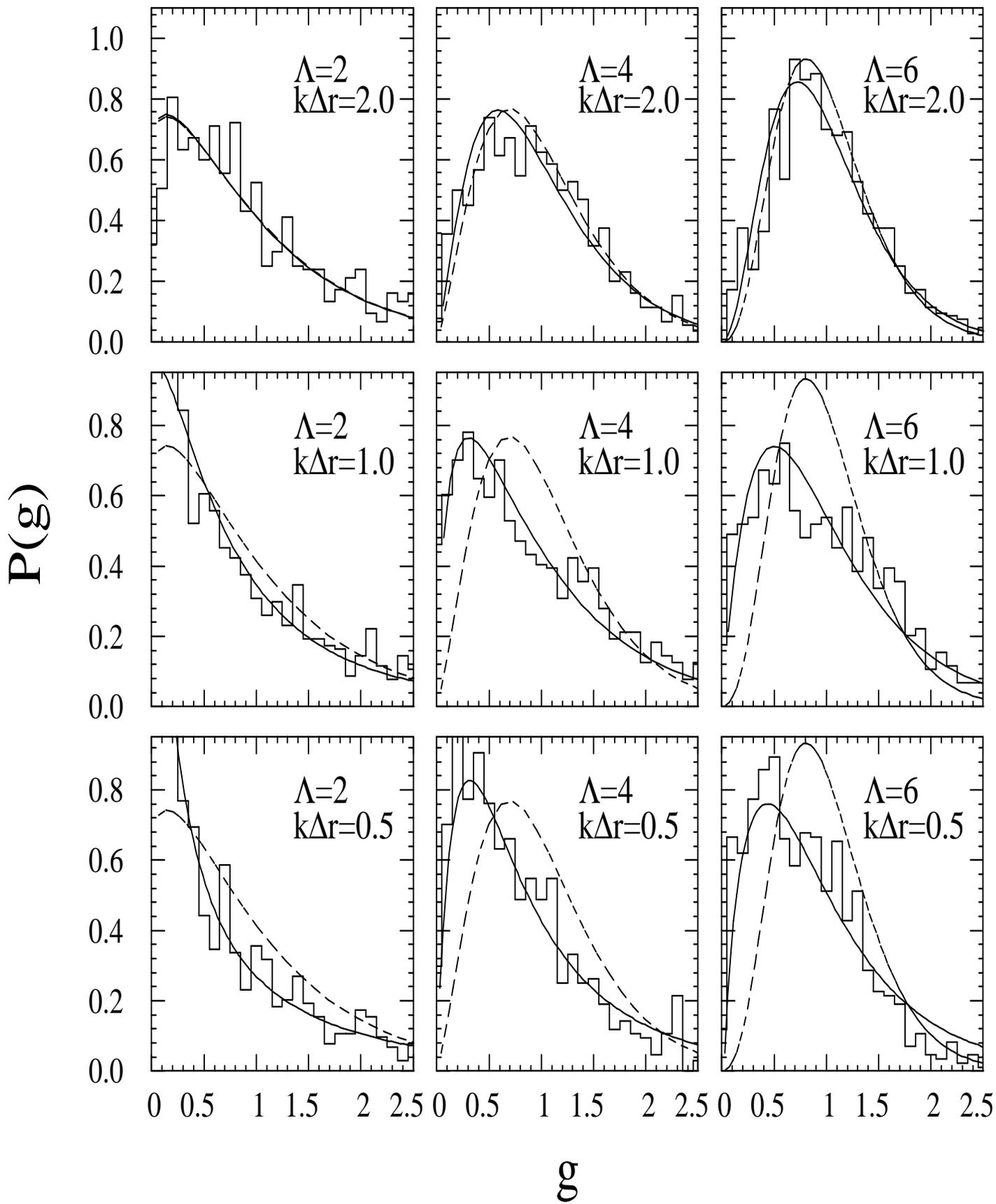
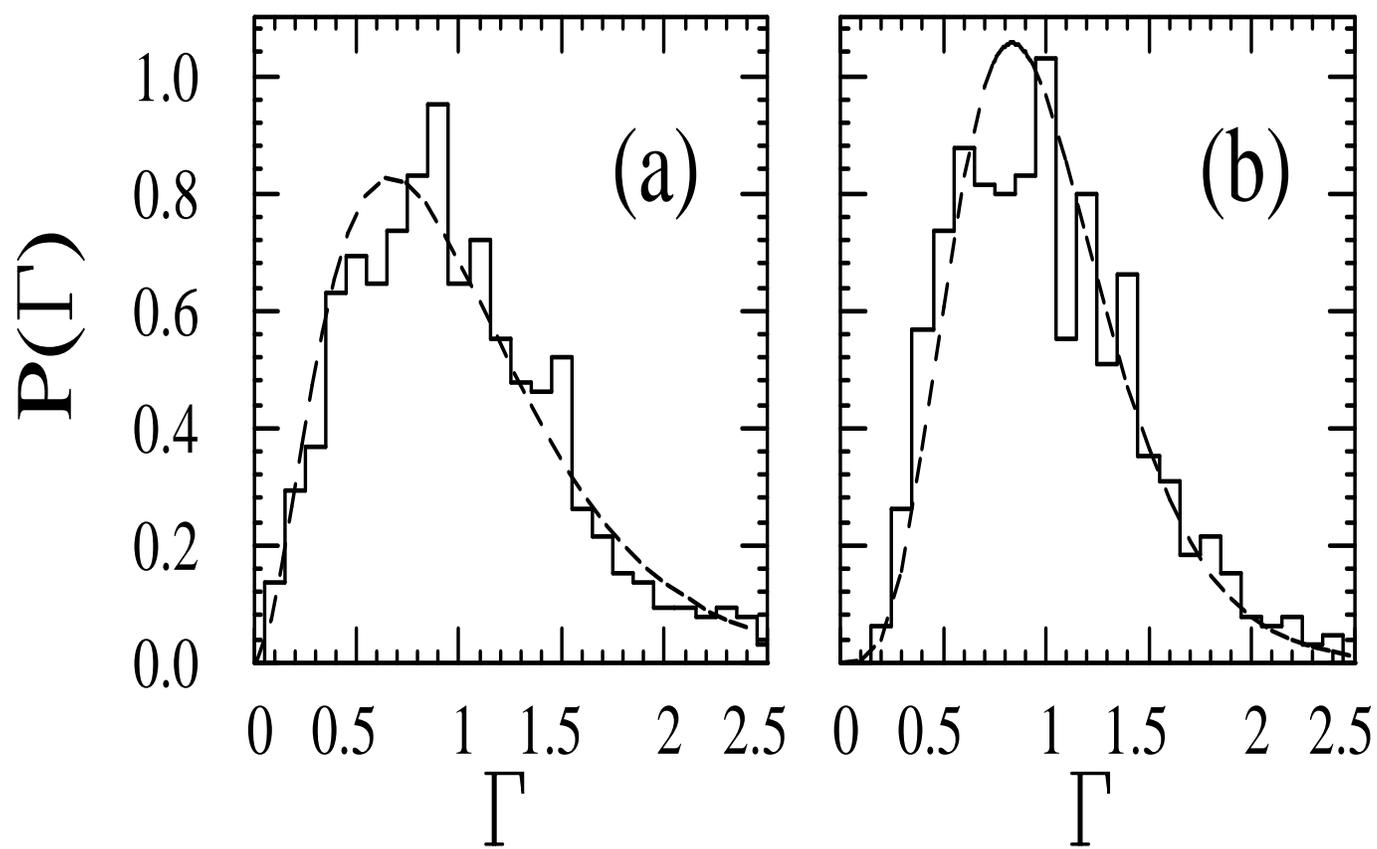
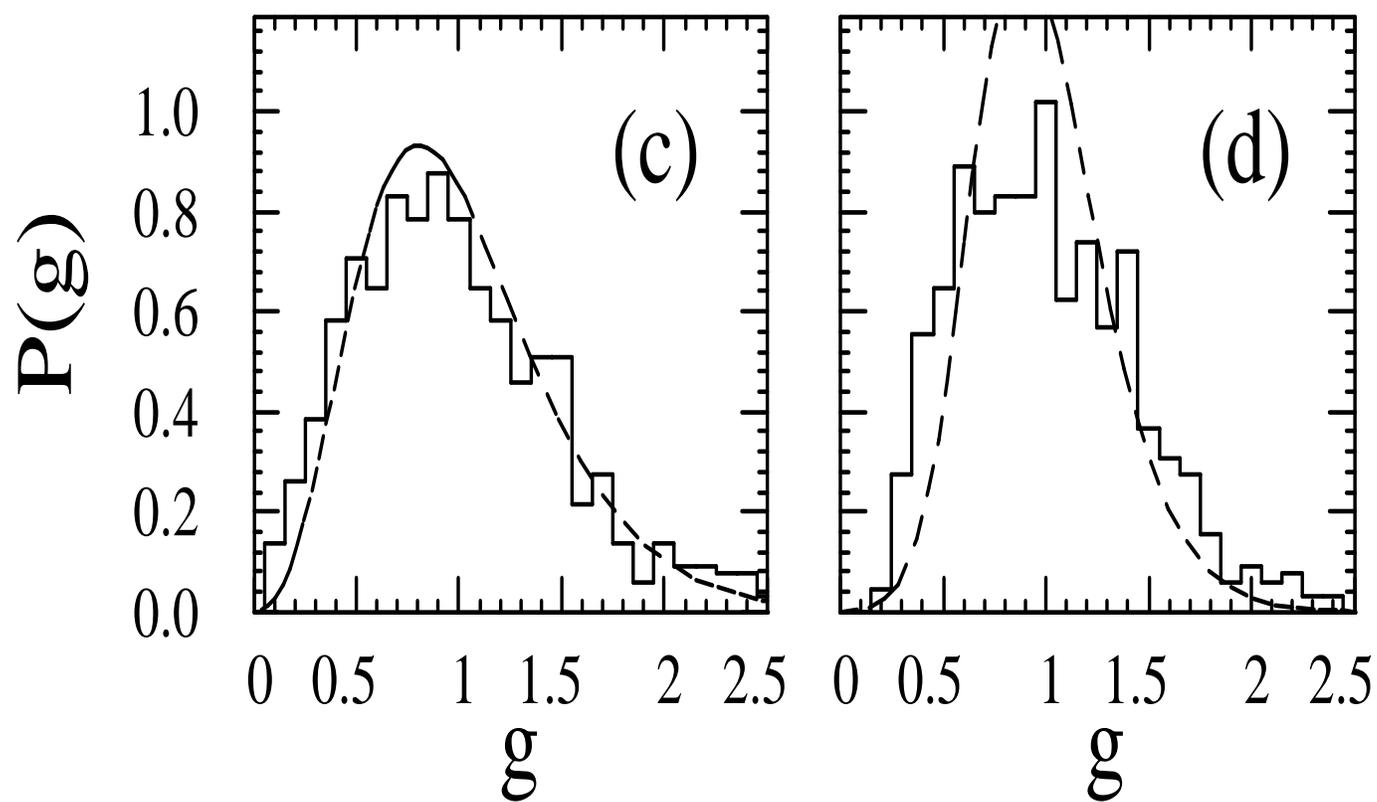
\end{document}